\documentclass[prb,aps,epsf,twocolumn,showpacs]{revtex4-1}
\usepackage{dcolumn}
\usepackage{bm}
\usepackage{epsfig}
\usepackage{latexsym}
\usepackage{amsmath}
\usepackage{amssymb}
\usepackage{array}
\usepackage{hyperref}
\usepackage{array}
\usepackage{cancel}
\usepackage{ulem}

\usepackage{xcolor,colortbl}
\usepackage{multirow}

\usepackage{booktabs}
\newcolumntype{C}[1]{>{\centering\arraybackslash}m{#1}}
\AtBeginDocument{
\heavyrulewidth=.08em
\lightrulewidth=.05em
\cmidrulewidth=.03em
\belowrulesep=.65ex
\belowbottomsep=0pt
\aboverulesep=.4ex
\abovetopsep=0pt
\cmidrulesep=\doublerulesep
\cmidrulekern=.5em
\defaultaddspace=.5em
}
\usepackage{booktabs}
\newcolumntype{R}[1]{>{\raggedleft\arraybackslash}p{#1}}
\AtBeginDocument{
\heavyrulewidth=.08em
\lightrulewidth=.05em
\cmidrulewidth=.03em
\belowrulesep=.65ex
\belowbottomsep=0pt
\aboverulesep=.4ex
\abovetopsep=0pt
\cmidrulesep=\doublerulesep
\cmidrulekern=.5em
\defaultaddspace=.5em
}

\newlength{\arrayrulewidthOriginal}

\usepackage{hhline}

\renewcommand{\>}{\rangle}

\renewcommand{\v}[1]{\mathbf{#1}} % \v -> vector (bf)

\newcommand{\Z}{\mathbb{Z}}

\newcommand{\header}[1]{\vspace{4pt}\noindent{\bf #1 --}}

\newcommand{\Ket}[1]{\left|#1  \right>}
\newcommand{\Bra}[1]{\left<#1  \right|}

\usepackage{pifont}% http://ctan.org/pkg/pifont
\newcommand{\cmark}{\ding{51}}%
\newcommand{\xmark}{\ding{55}}%

\renewcommand{\emph}[1]{{\it #1}}
\definecolor{DGreen}{rgb}{0, 0.5, 0.0}

\begin{document}
\title{Symmetry constraints on many-body localization}

\author{Andrew C. Potter$^1$}%\email{acpotter@berkeley.edu}
\author{Romain Vasseur$^{1,2}$}%\email{rvasseur@berkeley.edu}
\affiliation{$^1$Department of Physics, University of California, Berkeley, CA 94720, USA}
\affiliation{$^2$Materials Science Division, Lawrence Berkeley National Laboratories, Berkeley, CA 94720}
\begin{abstract}
We derive general constraints on the existence of many-body localized (MBL) phases in the presence of global symmetries, and show that MBL is not possible with symmetry groups that protect multiplets (e.g. all non-Abelian symmetry groups). Based on simple representation theoretic considerations, we derive general Mermin-Wagner-type principles governing the possible alternative fates of non-equilibrium dynamics in isolated, strongly disordered quantum systems. Our results rule out the existence of MBL symmetry protected topological phases with non-Abelian symmetry groups, as well as time-reversal symmetry protected electronic topological insulators, and in fact all fermion topological insulators and superconductors in the 10-fold way classification. Moreover, extending our arguments to systems with intrinsic topological order, we rule out MBL phases with non-Abelian anyons as well as certain classes of symmetry enriched topological orders.
\end{abstract}
\maketitle

%\header{Introduction} 
The concept of symmetry plays a crucial role in our understanding of phases of matter. The interplay of symmetry and dimensionality leads to very general constraints on possible types of symmetry breaking phases and phase transitions, such as the Peierls or Mermin-Wagner theorems. Going beyond Landau's theory of phases and phases transitions in terms of spontaneous symmetry breaking, it was recently understood that symmetries can protect topological distinctions among short-range entangled phases of matter  -- leading to the concept of symmetry protected topological (SPT) phases~\cite{PhysRevB.80.155131,PhysRevB.84.235128,PhysRevB.83.075102,PhysRevB.85.075125,PhysRevB.83.075103,Chen1604} exemplified by the celebrated electronic topological insulators (TI)~\cite{PhysRevLett.95.226801,Bernevig1757,PhysRevLett.98.106803,PhysRevB.75.121306,PhysRevB.79.195322,RevModPhys.82.3045} -- and that symmetry can also enrich the possibilities of quantum phases with long-range entanglement and intrinsic topological order~\cite{PhysRevB.65.165113,PhysRevLett.105.246809,PhysRevB.83.195139,PhysRevB.86.115131,PhysRevB.87.155115}. 

Whereas traditionally, the existence of phases and phase transitions is considered within the framework of equilibrium statistical mechanics, a sharp notion of quantum phases can be extended to a certain class of far-from-equilibrium quantum systems that fail to self-thermalize in isolation. Highly excited eigenstates in such {\it many-body localized} (MBL) systems~\cite{FleishmanAnderson,Gornyi,BAA,PhysRevB.75.155111,PalHuse} have properties akin to quantum groundstates~\cite{BauerNayak}, leading to the prospect of quantum coherent phenomena and universal dynamics~\cite{PhysRevLett.109.017202,PhysRevLett.110.260601,PhysRevLett.111.127201,PhysRevB.90.174202,BahriMBLSPT,PhysRevLett.113.147204,PhysRevB.91.140202}, and symmetry-breaking, topological or SPT quantum orders at high energy density~\cite{HuseMBLQuantumOrder,BauerNayak,PekkerRSRGX,PhysRevB.89.144201,BahriMBLSPT,2015arXiv150600592P,2015arXiv150505147S}. 

Given the fundamental role of symmetry in our understanding of equilibrium phases of matter, it is natural to expect very general symmetry principles to play a crucial role in MBL systems. While certain examples of symmetry based constraints on localization have been identified~\cite{QCGPRL,2015arXiv151004282V,2016arXiv160205194P}, a full and systematic understanding remains to be obtained. In this paper, we argue that the local conserved quantities that define MBL systems transform independently under the global symmetry, leading to an extensive number of local degeneracies in the presence of non-Abelian symmetries. This local action leads to very general constraints that dictate the fate of the excited state dynamics of strongly disordered systems on symmetry grounds alone. Namely, we show that symmetry preserving MBL phases are not possible with non-Abelian symmetries, either discrete or continuous. This eliminates out the possibility of SPT and symmetry enriched topological (SET) order protected by non-Abelian symmetries, and for instance means that it is impossible to localize spin-$\frac{1}{2}$ electrons with time-reversal symmetry, ruling out the realization of electronic TIs in MBL settings. Moreover, based on the representation theory of the symmetry group, we derive very general Mermin-Wagner type principles governing the stability and possible fates of strongly disordered systems with symmetry. For example, systems with continuous non-Abelian symmetries inevitably thermalize even at strong disorder, whereas systems with discrete non-Abelian symmetries may yield non-ergodic phases that are either well localized but with spontaneous symmetry breaking or are delocalized and quantum-critical.
%This strongly constraints the ability to protect quantum order at finite energy density using localization, and rules out the existence of MBL phases with non-Abelian symmetry. 
Our results also constraint the emergence of non-equilibrium quantum phases with intrinsic topological order and anyonic excitations, with the notion of anyonic fusion algebras replacing the group representation theory. For example, we rule out the possibility of MBL with non-Abelian anyonic excitations.
%with the notion of anyonhave important consequences for MBL-protected eigenstate SPT order, non-Abelian topological order, and anyonic edge modes. 

%\header{Many-body localization}
\section{Many-body localization and global symmetries}

\subsection{Many-body localization and local conserved quantities}

We begin by fixing a formal definition of MBL in terms of a complete set of quasi-local conservation laws~\cite{PhysRevLett.111.127201,PhysRevB.90.174202, VoskAltmanPRL13, 2014arXiv1403.7837I}, which has been widely adopted, and from which all of the known phenomenology of MBL systems such as area-law entanglement~\cite{BauerNayak}, absence of transport, slow dephasing and entanglement growth~\cite{PhysRevB.77.064426,PhysRevLett.109.017202,PhysRevLett.110.260601} directly follow. Specifically, we define MBL in terms of the existence of a complete set of conserved quantities, $\{n_\alpha\}$ ($\alpha = 1\dots N$), each of which takes values from a set $\lbrace 1, \dots, d_\alpha \rbrace$, with the associated quasi-local conserved projectors called l-bits~\cite{PhysRevLett.111.127201,PhysRevB.90.174202}
\begin{equation}
\Pi^{n_\alpha}_\alpha = \sum_{n_{\beta \neq \alpha}=1}^{d_\beta} \Ket{n_1 \dots n_\alpha \dots n_N } \Bra{n_1 \dots n_\alpha \dots n_N },
\end{equation}
that are exponentially well-localized within a localization length $\xi$ of position $\v{r}_\alpha$ and that commute with the Hamiltonian $\left[\Pi^{n_\alpha}_\alpha,H \right]=0$. We further assume that the quasi-local conserved quantities are related to local operators (with finite support) by dressing them with a quasi-local unitary transformation, which produces their exponential tails. This structure is obtained in all known examples of MBL, and in fact can be taken as the definition of the l-bits being quasi-local.

%to strictly local operators with finite range of support. 
%This assumption is analogous to characterizing, e.g. short-range entangled symmetry breaking equilibrium states of matter, as being smoothly connected to a ``fixed-point" wave-function  with zero-correlation length limit (e.g. a product state of spins pointing up for a ferromagnet).

Note that the number of conserved quantities $N$ can in general be different from the number of physical sites $L$. We will restrict our attention the case where all of the energy eigenstates (or quasi-energy eigenstates for Floquet systems~\cite{PhysRevLett.115.030402,PhysRevLett.114.140401}) are MBL, and neglect the theoretically more delicate case of a partially localized spectrum with a many-body mobility edge~\cite{PhysRevB.93.014203,2016arXiv160801328G}. In this case, the many-body eigenstates can be uniquely labeled by product states with definite values of the conserved quantities: $H|\{n_\alpha\}\> = E(\{n_\alpha\})|\{n_\alpha\}\>$, where $E$ is a quasi-local function of $n_\alpha$. In the following, we will investigate whether MBL can be compatible with additional exact degeneracies due to symmetry.

\subsection{Local symmetry action}
\label{subsecFactor}
Having defined MBL we now derive some general constraints on many-body localization in the presence of symmetry. To begin, we start by showing that symmetry acts locally on the l-bits. 

Consider a lattice of sites $i$ containing quantum degrees of freedom that transform under a (possibly reducible) representation, ${\mathcal V}$, of a symmetry group $G$ -- for example a chain of spins-$\frac{1}{2}$ $(\mathcal{V})$, with spin-rotation symmetry $G=SU(2)$. The Hilbert space of this system decomposes into a tensor product of on-site Hilbert spaces, ${\mathcal H}={\mathcal V}^{\otimes L}$. In order that different symmetry operations have non-trivial action on the physical degrees of freedom, we will demand that ${\mathcal V}$ is a faithful representation of the symmetry $G$, and that we cannot merely group degrees of freedom into larger clumps that transform under a simpler symmetry $G'$. The former condition implies that all irreducible representations (irreps) of $G$ are contained in the tensor product ${\mathcal V}^{\otimes n}$ for sufficiently large $n$.

We define a symmetry preserving MBL phase as one in which the local conserved quantities labelling is consistent with the symmetry $\left[ \Pi^{n_\alpha}_\alpha , \prod_i g_i\right]=0$, where $g_i \in {\cal V}$ is the representation of the symmetry generator $g\in G$ on the site $i$. This is equivalent to having each set of conserved quantities $\lbrace n_\alpha \rbrace$ label a multiplet of states ${\mathcal V}_{n_1,n_2,\dots,n_N} $ that form a representation of $G$. To construct the local action of the symmetry, let us proceed as follows. Let us assume that at least one of the eigenstates labelled by a given set ${\lbrace n^0_\alpha \rbrace}$ is non-degenerate and transforms trivially under the symmetry, so that the corresponding representation ${\mathcal V}_{n^0_1,n^0_2,\dots,n^0_N}={\mathcal V}_{0}$ is the trivial representation  (or singlet) with dimension 1. We then create a local excitation by changing the label $n^0_\alpha$ to $n_\alpha$ for a given l-bit $\alpha$: the resulting eigenstate(s) then transform in a different representation ${\mathcal V}_{n_\alpha} \equiv {\mathcal V}_{n^0_1,\dots,n_\alpha,\dots, n^0_N}$ which will generically be irreducible (otherwise, one may add generic local perturbations to reduce it). Because the change $n^0_\alpha \to n_\alpha$ is local, all the eigenstates $\Ket{n^0_1,\dots,n_\alpha,\dots, n^0_N; p}$ with $p = 1,\dots, {\rm dim} {\mathcal V}_{n_\alpha} $ in this representation and the singlet eigenstate $\Ket{\lbrace n^0_\alpha \rbrace}$ should differ only locally around $\v{r}_\alpha$. Let us now repeat the process to excite a different l-bit, $\beta$, and let ${\mathcal V}_{n_\alpha n_\beta}$ be the representation corresponding to the configuration where we changed the labels  $n_\alpha^0 \to n_\alpha$ and $n_\beta^0 \to n_\beta$ on two different locations $\alpha$ and $\beta$. 

We now show that the symmetry action factorizes on these two excitations. Since by their definition any quasi-local set of l-bits can obtained by ``dressing'' strictly local operators by a finite depth (quasilocal) unitary transformation, it suffices to consider the case of strictly local l-bits ({\it i.e.} with support only on a finite number of sites).
%these arguments also hold away from the strictly local limit.
%For strictly local l-bits ({\it i.e.} with support only on a finite number of contiguous sites), 
For this case, it is clear that if $\v{r}_\alpha$ and $\v{r}_\beta$ are sufficiently far apart so that the corresponding supports do not overlap, then the action of the symmetry factorizes on the two l-bits $\alpha$ and $\beta$: ${\mathcal V}_{n_\alpha n_\beta} = {\mathcal V}_{n_\alpha} \otimes {\mathcal V}_{n_\beta}$. Repeating the argument for an extensive number of l-bits $\alpha_1$, $\alpha_2$, \dots, $\alpha_p$ with $p \sim {\cal O} (N) \sim {\cal O}(L)$ sufficiently far apart so that their support do not overlap, we find that the symmetry action factorizes on the local l-bits ${\mathcal V}_{n_{\alpha_1} \dots n_{\alpha_p}} = {\mathcal V}_{n_{\alpha_1}} \otimes \dots \otimes {\mathcal V}_{n_{\alpha_p}}$. 

%Since we assume that any exponentially well-localized set of l-bits can be deformed to the ``fixed-point" limit of strictly local l-bits by a finite depth unitary transformation, these arguments also hold away from the strictly local limit.

\subsection{Examples of local symmetry action}This local factorization of the symmetry on the l-bits is particularly obvious for models of MBL paramagnets, such as the Ising paramagnet in arbitrary dimension with $G={\mathbb Z}_2$
\begin{equation}
H = - \sum_{i=1}^{L}  h_i \sigma_i^x + \dots
\label{eq:HIsing}
\end{equation}
where the dots represent small (but arbitrary) symmetry-preserving perturbations. 
The eigenstates of Eq.~\ref{eq:HIsing} are related to product states of definite $\sigma_i^x=\pm 1$ by a finite depth (quasi-local) unitary transformation, $U$, such that the local conserved quantities of this MBL systems are the dressed projectors $\Pi_{\alpha_i}^{n=0,1} = U^\dagger \frac{1 \pm \sigma_i^x}{2} U$.
%The transverse field term $\sigma_i^x = \pm 1$ simply gives different energies to the even and odd sectors of the ${\mathbb Z}_2$ symmetry generated by $\prod_i \sigma_i^x$. The local conserved quantities of this MBL system are the dressed projectors $\Pi_{\alpha_i}^{n=0,1} = U^\dagger \frac{1 \pm \sigma_i^x}{2} U$ where $U$ is a finite depth (quasilocal) unitary transformation.
In this case, the local action of the symmetry is simply given by 
\begin{equation}
\hat{g}_{\alpha_i} = U^\dagger \sigma^x_i U,  
 \end{equation}
with $\left[\hat{g}_{\alpha_i} , H \right] =0$, since $\hat{g}_{\alpha_i} $ commutes with all the conserved quantities $ \Pi_{\alpha_j}^n$, readily verifying %We therefore see clearly in this case
 that the global symmetry $\left[\prod_i g_i, H\right]=0$ with $g_i=\sigma_i^x$ is promoted to a local symmetry $\left[\hat{g}_{\alpha_i} , H \right] =0$ for the MBL system.

This construction can be readily generalized to the generic MBL paramagnet Hamiltonian $H_{\rm para} = - \sum_i \sum_n h^n_{i} P_i^n + \dots$ where the $P_i^n$'s are projection operators onto the different irreps (``channels'') in the decomposition ${\mathcal V} = \oplus_n {\mathcal V}_n$ ($\sum_n P_i^n = 1$) of the on-site representation ${\mathcal V}$ of $G$, and the dots represent generic weak perturbations. As in the Ising example, the local conserved quantities of this MBL system are the dressed projectors $\Pi_{\alpha_i}^n = U^\dagger P_i^n U$ where $U$ is a finite depth (quasilocal) unitary transformation, and the local action of the symmetry is simply given by  $\hat{g}_{\alpha_i} = U^\dagger g_i U$, where $g_i$ in the representation of the group element $g$ on site $i$. 

A less straightforward example are SPT phases, which cannot be \emph{continuously} connected to a trivial paramagnet while preserving symmetry. However, the SPT eigenstates are \emph{non-continuously} deformable, via a finite-depth unitary transformation $U_\text{SPT}$ that preserves the symmetry everywhere in the bulk, to a trivial paramagnet (see e.g.~\cite{PhysRevB.87.155114}, and Appendix~\ref{AppA} for a specific example). We can then utilize the construction for paramagnets to identify, the local action of symmetry as being generated by $\hat{g}_{\alpha,i}=U^\dagger U_\text{SPT}^\dagger g_i U_\text{SPT}U$, where $U$ is the quasi-local unitary that dresses the l-bits. These generators form an ordinary local representation of symmetry in the bulk, but act non-trivially (e.g. projectively in 1D) at the edges of the system.

\section{MBL and non-Abelian symmetry}

 The local factorization of the symmetry on the l-bits has important consequences when $G$ is non-Abelian, for which some irreps are necessarily multidimensional. Intuitively, this signals an obstacle to localization, since a generic MBL state will contain many of these multidimensional excitations, each with local degrees of freedom that cost no energy to excite, and can therefore freely inter-resonate with each other leading to a breakdown of localization. This rules out the existence of MBL paramagnets with Potts (permutation group $G=S_n$) or non-chiral clock (Dihedral group $G=D_n = {\mathbb Z}_n \rtimes  {\mathbb Z}_2 $) symmetry for instance. We emphasize that while our argument relied on the local integrability picture of MBL systems, we expect the main idea to be fairly general so that it would also rule out tentative MBL phases without an l-bit description (see discussion below).

%In an MBL phase, one would inevitably encounter a finite density of excitations that each have local degeneracies equal to the dimension of their corresponding irrep, producing a massively degenerate energy spectrum.
%faithful irreps have to be multidimensional. 
%\ACP{This immediately implies that, if one were to have an MBL phase in such a system, one would inevitably encounter a finite density of excitations that would each transform independently under one of these multidmensional irreps. Such excitations would each have a local degeneracy (ala Kramers degeneracy) equal to the dimension of their respective irrep, leading to an exponential-in-system-size degeneracy of the energy spectrum.
% Since a finite density of the irreps ${\mathcal V}_{n_{\alpha_k}}$, $k=1,\dots,p$ in the decomposition ${\mathcal V}_{n_{\alpha_1} \dots n_{\alpha_p}} = {\mathcal V}_{n_{\alpha_1}} \otimes \dots \otimes {\mathcal V}_{n_{\alpha_p}}$ should be faithful and therefore multidimensional, this implies directly that the equal-energy eigenstates transforming in the representation ${\mathcal V}_{n_{i_1} \dots n_{i_p}}$ are exponentially degenerate. 
%Such a huge degeneracy cannot be stable to \ACP{generic perturbations, no matter how small,} \sout{quantum fluctuations,}  so that MBL cannot exist as a stable non-equilibrium phase in the presence of non-Abelian symmetry.  \sout{(assuming that all MBL systems admit a zero-correlation length limit where the l-bits have a finite support).}  }

More formally, if we were to have an MBL system with non-Abelian symmetry $G$, then the local conserved quantities would transform as irreps of the symmetry group $G$ so that each l-bit, $n_\alpha$, labels an irrep ${\mathcal V}_{n_\alpha}$ of $G$. The Hilbert space therefore has a symmetry preserving tensor structure in the l-bit space ${\mathcal H}=\otimes_\alpha {\mathcal V}_\alpha$,  where the representation ${\mathcal V}_\alpha$ is reducible and can be decomposed as ${\mathcal V}_\alpha = \oplus_{n_\alpha=1}^{d_\alpha} {\mathcal V}_{n_\alpha}$. Since the physical degrees of freedom transform in a faithful representation ${\mathcal V}$ of the symmetry, at least a finite density of the ${\mathcal V}_\alpha$'s should be faithful representations of $G$ as well. If $G$ is non-Abelian, this immediately implies that some irreps ${\mathcal V}_{n_\alpha}$ should have dimension larger than 1 so that the quantum numbers $n_\alpha$ must be supplemented with an additional number $p_\alpha=1, \dots, D_{n_\alpha} = {\rm dim}  {\mathcal V}_{n_\alpha}  $ to label uniquely an eigenstate. This finite density of local multidimensional irreps leads to an exponential degeneracy of the eigenstates of $H$ since the energy cannot depend on the extra labels $p_\alpha$. In other words, the global symmetry $G$ is promoted to a local symmetry 
\begin{equation}
[{\hat g}_{n_\alpha}, \Pi_\beta^{n_\beta}] = [{\hat g}_{n_\alpha}, H] =0,
\end{equation}
because of the many-body localized structure of the eigenstates, with ${\hat g}_{n_\alpha}$ being the representation of $g \in G$ in ${\mathcal V}_{n_\alpha} $, acting locally around position $\v{r}_\alpha$. This extended local symmetry leads to local degeneracies if there are multidimensional irreps, leading in turn to a massive  exponential-in-system-size degeneracy of all eigenstates. Such degenerate eigenstates are inherently unstable, even to infinitesimally small, perturbations. However, the crucial point is that there is no local and symmetry-preserving way to resolve this degeneracy. Hence, either the symmetry or the localization must break down. Which of these fates may occur depends on the group structure, and below, we will identify some simple governing principles based on the number and dimensions of the irreps of the symmetry group. 

Note that whereas the above discussion assumed a full factorization of the symmetry on the l-bits for simplicity (so that the Hilbert space factorizes as ${\mathcal H}=\otimes_\alpha {\mathcal V}_\alpha$), the existence of exponentially-degenerate eigenstates only requires a partial factorization of the symmetry for a general excitation involving an extensive number of l-bits far-enough apart  ${\mathcal V}_{n_{\alpha_1} \dots n_{\alpha_p}} = {\mathcal V}_{n_{\alpha_1}} \otimes \dots \otimes {\mathcal V}_{n_{\alpha_p}}$, which we showed in Sec.~\ref{subsecFactor}. 

%This is inherently unstable to perturbations, which will lead to a break down of MBL.

\renewcommand{\arraystretch}{1.3}
\newcolumntype{X}{>{\centering\arraybackslash}p{1in}}
\newcolumntype{Y}{>{\centering\arraybackslash}m{2in}}
\newcolumntype{Z}{>{\centering\arraybackslash}m{2in}}
\begin{table*}
\begin{tabular}{  X  Y  Z  }
\toprule[1pt]
   $G$ & $\#~{\rm irreps}<\infty$  & $\#~{\rm irreps}=\infty$  \\
  
\midrule[1pt]

  $\left| {\cal V} \right| = 1$  & MBL\textcolor{DGreen}{\cmark}   Ex: ${\mathbb Z}_n$ & MBL\textcolor{DGreen}{\cmark}   Ex: $U(1)$\\
%    \hhline{|=|==|==|}
\midrule[0.2pt]
  \multirow{2}{*}{$1 <\left| {\cal V} \right| < \infty$} & MBL\textcolor{red}{\xmark}  Ex: ${\mathbb Z}_n \rtimes  {\mathbb Z}_2 $ & MBL\textcolor{red}{\xmark}   Ex: $U(1) \rtimes {\mathbb Z}_2$\\ 
 %\cmidrule(lr){2-2}
    %\cmidrule(lr){3-3}% \cline{2-5}
         &$\rightarrow$ MBL+SSB~(or QCG?) &  $\rightarrow$ MBL+SSB\\
%          &$\rightarrow$ QCG (?) & $\rm{ }$ \\
\midrule[0.2pt]
  \multirow{2}{*}{$\left|{\cal V} \right|= \infty$}  & \multirow{2}{*}{N/A}  & MBL\textcolor{red}{\xmark}   Ex: $SU(2)$\\  %\cline{4-5}
     & &  $\rightarrow$ Thermalization only\\
     \bottomrule[1pt]

\end{tabular}

\caption{{\bf Symmetry constraints on MBL:} Possible phases of an isolated interacting system at strong disorder in terms of the representation theory of its symmetry group $G$. The relevant parameters are the number of irreps and the dimension of the largest irrep $\left| {\cal V} \right| = {\rm sup}_k \lbrace {\rm dim} {\cal V}_k \rbrace$. If $G$ is Abelian ($\left| {\cal V} \right| $=1), then a many-body localized phase is possible at strong disorder. If on the other hand $G$ is non-Abelian, a symmetry-preserving MBL phase is not allowed, giving rise to either thermalization, MBL with the symmetry spontaneously broken (SSB) to an Abelian subgroup, or to non-trivial quantum critical glasses (QCG) depending on the properties of the symmetry group (see text). }
\label{TabSym}
\end{table*}

\section{General symmetry principles} 
%As we have argued above, non-Abelian symmetries in an MBL system would lead to an extensive number of local degeneracies which in turn would imply exponentially degenerate eigenstates. Such huge degeneracies should be lifted by arbitrarily small perturbations and therefore cannot be generic, and several possible outcomes are conceivable. 
Above, we have shown that non-Abelian symmetries are not consistent with MBL phases, and we now seek some general insight into the possible fates of an isolated non-equilibrium system with non-Abelian symmetries. If disorder is too weak, we expect that the putative local degenerate excitations
will strongly overlap and inter-resonate, 
%so that quantum fluctuations naturally lift the degeneracies by 
driving the system into a thermalizing phase. Hence, we will subsequently focus on the regime of strong disorder, considering various classes of non-Abelian symmetry groups in turn.

When the non-Abelian group $G$ has irreps of bounded dimension, with either infinitely many irreps (as for the group $G= U(1) \rtimes  {\mathbb Z}_2 $ where all irreps have dimension $ \leq2$) or a finite number of them (as for any finite non-Abelian symmetry), there are a few options to lift the degeneracies at strong disorder. One possible outcome is that the system forms an MBL state in which symmetry is spontaneously broken  down to an Abelian subgroup by choosing a particular set of the numbers $p_\alpha$. This spontaneous symmetry breaking (SSB) scenario was  previously demonstrated for the particular example of a random XXZ spin chain, equivalent to fermions with particle-hole symmetric disorder with symmetry group $G= U(1) \rtimes  {\mathbb Z}_2 $~\cite{2015arXiv151004282V} using renormalization group techniques and numerical simulations. Another possible option for finite groups would be for the system to form a symmetry preserving ``quantum critical glass'' (QCG) which is neither thermal nor exponentially localized, and that cannot be described in terms of independent conserved quantities (examples of such phases have been uncovered in analytically solvable random anyonic chains~\cite{QCGPRL}).  It would be very interesting to find a concrete example of such a QCG phase in a random spin chain with non-Abelian symmetry. Note however that our argument also rules out marginal or quantum critical MBL states~\cite{VoskAltmanPRL13, PekkerRSRGX, PhysRevLett.112.217204, NandkishorePotterScaling} where independent l-bits exist, just with algebraic rather than exponential tails
%the conserved quantities have a structure very similar to the MBL case with algebraic instead of exponential tails 
(as for the critical random transverse field Ising chain for example~\cite{PekkerRSRGX, PhysRevLett.112.217204,PhysRevB.93.104205}).

If the non-Abelian symmetry group $G$ is continuous, e.g. $G=SU(2)$, then it will possess infinitely many irreps with arbitrarily large dimension. E.g., for $G=SU(2)$, irreps are labelled by spin size $S$ and have dimension $2S+1$ where $S$ can be arbitrarily large. Then following Ref.~\onlinecite{QCGPRL} in a large many-body system, one will encounter excitations with arbitrarily large local degeneracy, $D$ (large ``spins"), whose quantum fluctuations are suppressed as $1/D$, leading to effectively classical dynamics, and resulting in thermalization, even for arbitrarily strong disorder~\cite{PhysRevB.80.115104}.
% it is impossible to localize the arbitrarily large irreps  (large ``classical spins'') even at arbitrarily strong disorder, leading inevitably to a classical thermal phase. 
Note that it is furthermore not possible to realize an MBL phase by spontaneously breaking the continuous non-Abelian symmetry, as this would produce a delocalized Goldstone mode that would act as a bath~\cite{PhysRevB.27.5592,PhysRevB.68.134207,NandkishorePotterScaling,PhysRevLett.116.116601}, so that thermalization is the only possible scenario.
 
The only scenario that permits stable MBL phases with symmetry are Abelian groups, whose irreps all have dimension $1$, avoiding the pitfalls of the above examples. These different scenarios are summarized in Tab.~\ref{TabSym}.

\section{Consequences for non-equilibrium topological phases}

\subsection{Consequences for SPT order} 
The above-identified obstruction to MBL rules out the possibility of localization stabilized SPT order (or Floquet SPT order~\cite{PhysRevB.92.125107,2016arXiv160202157V,2016arXiv160206949V,2016arXiv160204804E,2016arXiv160205194P}) with non-Abelian symmetry groups such as the Haldane chains with continuous $SO(3)$ symmetry~\cite{PhysRevLett.59.799} -- as these phases require both symmetry and MBL to occur at high energy density. This further constrains the many-body localizability of SPT phases~\cite{2015arXiv150600592P,2015arXiv150505147S}.

%The explicit form of the local action of the symmetry is slightly more complicated since the model Hamiltonians for SPT phases necessarily involve projectors acting on multiple sites and overlapping with each other. However, we can use the fact that SPTs are short-range entangled state that can be deformed in a quasi-local way into a symmetry-preserving trivial product state $H_{\rm SPT} = U_{\rm SPT}^\dagger H_{\rm para} U_{\rm SPT}$,  where $U_{\rm SPT}$ is a local unitary that preserves the symmetry in the bulk (see {\it e.g.}~\cite{PhysRevB.87.155114}). If $\prod_i g_i$ commutes with $H_{\rm para}$, then $\prod_i U_{\rm SPT}^\dagger  g_i U_{\rm SPT}$ is a symmetry of $H_{\rm SPT}$ so we can use our results for MBL paramagnets: the local action of the symmetry commuting with the MBL Hamiltonian is $\hat{g}_{\alpha_i} = U^\dagger U_{\rm SPT}^\dagger g_i U_{\rm SPT} U$,  where $U$ is the quasi-local unitary that dresses the l-bits.}

%This construction can be readily extended to the commuting projector Hamiltonians of interacting bosonic SPTs obtained from group cohomology~\cite{PhysRevB.87.155114}

 We remark that these results apply also to anti-unitary symmetries such as time reversal symmetry (TRS). The notion of local action of TRS is in general somewhat subtle, due to the nominally global action of complex conjugation. However, for MBL states, which by definition permit a tensor product state description, one may readily construct a well-defined local action of time-reversal~\cite{PhysRevB.86.115131,PhysRevX.5.041034}. A notable case, is that of spin-1/2 electrons with time-reversal symmetry. In a putative TRS MBL state of such particles, electronic excitations would exhibit a local two-fold Kramers degeneracy, spoiling the stability of the localized phase. In particular, this rules out the possibility of 2D and 3D TRS electron topological insulators~\cite{RevModPhys.82.3045} in MBL systems.

In fact, this and related obstructions rule out the possibility of physically realizing \emph{any} fermionic topological insulator in physically accessible dimensions ($d\leq 3$) in the 10-fold way classification~\cite{Kitaev10Fold,3149481}, for the following reasons. First, any of the topological superconducting classes require a pair condensate, which in ultra-cold atomic systems in which MBL may be realized~\cite{Schreiber842,2015arXiv150807026S,PhysRevLett.116.140401,2016arXiv160404178C}, implies the existence of a superfluid Goldstone mode which will lead to thermalization~\cite{PhysRevB.27.5592,PhysRevB.68.134207,NandkishorePotterScaling,PhysRevLett.116.116601}. Next, any non-superconducting TI class has either Kramers doublet fermions ($\mathcal{T}^2=-1$), a particle hole symmetry (leading to non-Abelian group structure)~\cite{2015arXiv151004282V}, 
%$\mathcal{C}$, that leads to group structure $U(1)\rtimes \mathcal{C}$ (for unitary particle hole symmetry, $\mathcal{C}$) or $U(1) \times \mathcal{S}$ (for antiunitary particle-hole symmetry, $\mathcal{S}$), both of which have dimension-two irreps and cannot be localized while preserving symmetry, 
or chiral edge states~\cite{NandkishorePotterScaling,2015arXiv150600592P} -- any of which prevent symmetry-preserving MBL. Whether any fermion SPT outside the 10-fold way is suitable for MBL protection can be examined on a case-by-case basis using the above criteria.

\subsection{Localization of Anyons} Symmetry can also lead to new topologically ordered phases and our results immediately imply that such symmetry enriched topological (SET) phases with non-Abelian symmetry cannot be many-body localized. Moreover, our arguments also rule out MBL protection of classes of SET order in which the global symmetry group is Abelian, but where the local symmetry action on fractionalized anyons is projective (requiring multidimensional local degeneracy)  and hence acts like a non-Abelian symmetry~\cite{PhysRevB.87.155115,PhysRevB.93.155121,Barkeshli:2014cna}. Examples of this class of phases include discrete gauge theories in which the electric domain walls are decorated by one-dimensional SPTs, such that the electric charge excitations transform as the ends of 1D SPTs, and hence have symmetry protected degeneracy that prevents symmetry preserving MBL of generic excited states in which such excitations are present at finite density at random locations.

Even without any additional global symmetry, our argument can be naturally generalized to topologically ordered systems in 2+1 dimensions with non-Abelian anyonic excitations. If such systems could be many-body localized, the finite density of exponentially localized non-Abelian anyons in generic eigenstates would lead to an exponential degeneracy of eigenstates (the quantum dimension of the anyons playing the role of the dimension of the irreps in our previous discussion). This forbids area-law entangled MBL phases with non-Abelian topological order, and simply reflects the fact that the topological Hilbert space of non-Abelian anyons does not have a local tensor product structure and that the notion of topological charge cannot be made local. In general, we expect interacting anyons to either thermalize or to form more exotic non-ergodic states that cannot be described in terms of independent l-bits, such as the QCG phase in 1D~\cite{QCGPRL}. 

%Consequences for the localizability of anyonic edge modes in one dimensional setups~\cite{PhysRevLett.100.096407,PhysRevB.79.161408,PhysRevB.82.041405,PhysRevX.2.031013,PhysRevX.2.041002,PhysRevB.86.195126,CAKpara,PhysRevX.4.011036} and their relation to universal topological quantum computation are discussed in in the supplemental material~\cite{SupMat}. 

\subsection{Localizability of anyonic edge modes}
\label{AppB}

The constraints on the localization of anyons discussed above also have consequences for one-dimensional ``trenches'' of non-Abelian anyons $\psi$ with quantum dimension $d_\psi$, such as the 1D chain of Majorana bound states ($d_\psi=\sqrt{2}$) that emerges from gapping out the edge of a 2D TI or fractional TI by proximity to alternating ferromagnetic and superconducting regions~\cite{PhysRevB.79.161408,PhysRevB.82.041405,PhysRevX.2.031013,PhysRevX.2.041002,PhysRevLett.100.096407,PhysRevB.79.161408,PhysRevB.86.195126,CAKpara,PhysRevX.4.011036}. Focusing on the topological low-energy (in-gap) sector, we can ask whether the 1D topological phase with anyonic edge modes obtained by dimerizing the couplings can be protected to finite energy density (within the topological sector) using MBL. In the perfectly dimerized limit, the eigenstates of such a system consist of two dangling anyonic edge modes and of anyonic excitations resulting from the fusion $\psi \times \psi $ on the bulk dimerized bonds. Our discussion implies that an MBL phase away from the perfectly dimerized limit can exist if and only if the anyons appearing when fusing $\psi$ with itself all have dimension one, which can occur only if $d_\psi^2=p$ is an integer --- corresponding to the so-called parafermionic zero modes~\cite{1742-5468-2012-11-P11020,PhysRevB.86.195126,CAKpara,PhysRevX.4.011036,AFpara} that generalize Majorana fermions ($p=2$). Only such parafermionic edge modes can be protected by MBL, which while interesting for topological quantum computing applications~\cite{RevModPhys.80.1083}, are not enough to realize a set of universal quantum gates. One-dimensional chains of anyons~\cite{PhysRevLett.98.160409, PhysRevLett.101.050401} whose braiding would provide a universal gate set  (such as Fibonacci anyons for example) cannot be many-body localized even by strongly dimerizing the couplings, and instead generically thermalize (at weak disorder), or form a non-ergodic QCG phase (at strong disorder) consistent with recent real-space renormalization group results~\cite{QCGPRL}.

\section{Discussion and generalizations}

In this paper, we showed that MBL is not possible with symmetry groups that protect degeneracies ({\it i.e.} that have multidimensional irreps). This ``no-go theorem'' relies on a specific definition of MBL in terms of local integrability~\cite{PhysRevLett.111.127201,PhysRevB.90.174202}, and the existence of a complete set of local conserved quantities (``l-bits'') so that all eigenstates are smoothly connected by a quasilocal unitary transformation to a zero correlation length limit. This l-bit picture  has become central to our current understanding of area-law entangled MBL phases, and underlies all of the phenomenology of MBL systems (absence of transport, logarithmic dephasing {\it etc.}). The existence of MBL phases beyond the l-bit picture is controversial, and could include systems with many-body mobility edges for instance~\cite{PhysRevB.93.014203,2016arXiv160801328G}.  We emphasize here that our argument can be naturally extended beyond the local integrability picture, and would rule out tentative MBL phases that would escape the l-bit description. The key point is that our argument relies only on the existence of local excitations over a symmetric eigenstate (say, the groundstate) that transform nicely under the symmetry. Even if the l-bit picture breaks down, excitations of an MBL system should be local, and we expect that they should naturally transform under irreps of the symmetry group. A finite density of such local excitations transforming according to irreps of dimension larger than one would immediately lead  to exponentially degenerate eigenstates, which are inherently unstable. While this argument can be made essentially rigorous within the l-bit picture, we expect the main idea to be fairly general so that it would rule out MBL phases without an l-bit description as well (if such phases do exist).

\header{Acknowledgments.} We thank T. Morimoto, S. Parameswaran and A. Vishwanath for insightful discussions. This work was supported by the Gordon and Betty Moore Foundation's EPiQS Initiative through Grant GBMF4307 (ACP) and the Quantum Materials Program at LBNL (RV).

\ \

\bibliography{MBL}

\begin{thebibliography}{81}%
\makeatletter
\providecommand \@ifxundefined [1]{%
 \@ifx{#1\undefined}
}%
\providecommand \@ifnum [1]{%
 \ifnum #1\expandafter \@firstoftwo
 \else \expandafter \@secondoftwo
 \fi
}%
\providecommand \@ifx [1]{%
 \ifx #1\expandafter \@firstoftwo
 \else \expandafter \@secondoftwo
 \fi
}%
\providecommand \natexlab [1]{#1}%
\providecommand \enquote  [1]{``#1''}%
\providecommand \bibnamefont  [1]{#1}%
\providecommand \bibfnamefont [1]{#1}%
\providecommand \citenamefont [1]{#1}%
\providecommand \href@noop [0]{\@secondoftwo}%
\providecommand \href [0]{\begingroup \@sanitize@url \@href}%
\providecommand \@href[1]{\@@startlink{#1}\@@href}%
\providecommand \@@href[1]{\endgroup#1\@@endlink}%
\providecommand \@sanitize@url [0]{\catcode `\\12\catcode `\$12\catcode
  `\&12\catcode `\#12\catcode `\^12\catcode `\_12\catcode `\%12\relax}%
\providecommand \@@startlink[1]{}%
\providecommand \@@endlink[0]{}%
\providecommand \url  [0]{\begingroup\@sanitize@url \@url }%
\providecommand \@url [1]{\endgroup\@href {#1}{\urlprefix }}%
\providecommand \urlprefix  [0]{URL }%
\providecommand \Eprint [0]{\href }%
\providecommand \doibase [0]{http://dx.doi.org/}%
\providecommand \selectlanguage [0]{\@gobble}%
\providecommand \bibinfo  [0]{\@secondoftwo}%
\providecommand \bibfield  [0]{\@secondoftwo}%
\providecommand \translation [1]{[#1]}%
\providecommand \BibitemOpen [0]{}%
\providecommand \bibitemStop [0]{}%
\providecommand \bibitemNoStop [0]{.\EOS\space}%
\providecommand \EOS [0]{\spacefactor3000\relax}%
\providecommand \BibitemShut  [1]{\csname bibitem#1\endcsname}%
\let\auto@bib@innerbib\@empty
%</preamble>
\bibitem [{\citenamefont {Gu}\ and\ \citenamefont
  {Wen}(2009)}]{PhysRevB.80.155131}%
  \BibitemOpen
  \bibfield  {author} {\bibinfo {author} {\bibfnamefont {Z.-C.}\ \bibnamefont
  {Gu}}\ and\ \bibinfo {author} {\bibfnamefont {X.-G.}\ \bibnamefont {Wen}},\
  }\href {\doibase 10.1103/PhysRevB.80.155131} {\bibfield  {journal} {\bibinfo
  {journal} {Phys. Rev. B}\ }\textbf {\bibinfo {volume} {80}},\ \bibinfo
  {pages} {155131} (\bibinfo {year} {2009})}\BibitemShut {NoStop}%
\bibitem [{\citenamefont {Chen}\ \emph {et~al.}(2011)\citenamefont {Chen},
  \citenamefont {Gu},\ and\ \citenamefont {Wen}}]{PhysRevB.84.235128}%
  \BibitemOpen
  \bibfield  {author} {\bibinfo {author} {\bibfnamefont {X.}~\bibnamefont
  {Chen}}, \bibinfo {author} {\bibfnamefont {Z.-C.}\ \bibnamefont {Gu}}, \ and\
  \bibinfo {author} {\bibfnamefont {X.-G.}\ \bibnamefont {Wen}},\ }\href
  {\doibase 10.1103/PhysRevB.84.235128} {\bibfield  {journal} {\bibinfo
  {journal} {Phys. Rev. B}\ }\textbf {\bibinfo {volume} {84}},\ \bibinfo
  {pages} {235128} (\bibinfo {year} {2011})}\BibitemShut {NoStop}%
\bibitem [{\citenamefont {Turner}\ \emph {et~al.}(2011)\citenamefont {Turner},
  \citenamefont {Pollmann},\ and\ \citenamefont {Berg}}]{PhysRevB.83.075102}%
  \BibitemOpen
  \bibfield  {author} {\bibinfo {author} {\bibfnamefont {A.~M.}\ \bibnamefont
  {Turner}}, \bibinfo {author} {\bibfnamefont {F.}~\bibnamefont {Pollmann}}, \
  and\ \bibinfo {author} {\bibfnamefont {E.}~\bibnamefont {Berg}},\ }\href
  {\doibase 10.1103/PhysRevB.83.075102} {\bibfield  {journal} {\bibinfo
  {journal} {Phys. Rev. B}\ }\textbf {\bibinfo {volume} {83}},\ \bibinfo
  {pages} {075102} (\bibinfo {year} {2011})}\BibitemShut {NoStop}%
\bibitem [{\citenamefont {Pollmann}\ \emph {et~al.}(2012)\citenamefont
  {Pollmann}, \citenamefont {Berg}, \citenamefont {Turner},\ and\ \citenamefont
  {Oshikawa}}]{PhysRevB.85.075125}%
  \BibitemOpen
  \bibfield  {author} {\bibinfo {author} {\bibfnamefont {F.}~\bibnamefont
  {Pollmann}}, \bibinfo {author} {\bibfnamefont {E.}~\bibnamefont {Berg}},
  \bibinfo {author} {\bibfnamefont {A.~M.}\ \bibnamefont {Turner}}, \ and\
  \bibinfo {author} {\bibfnamefont {M.}~\bibnamefont {Oshikawa}},\ }\href
  {\doibase 10.1103/PhysRevB.85.075125} {\bibfield  {journal} {\bibinfo
  {journal} {Phys. Rev. B}\ }\textbf {\bibinfo {volume} {85}},\ \bibinfo
  {pages} {075125} (\bibinfo {year} {2012})}\BibitemShut {NoStop}%
\bibitem [{\citenamefont {Fidkowski}\ and\ \citenamefont
  {Kitaev}(2011)}]{PhysRevB.83.075103}%
  \BibitemOpen
  \bibfield  {author} {\bibinfo {author} {\bibfnamefont {L.}~\bibnamefont
  {Fidkowski}}\ and\ \bibinfo {author} {\bibfnamefont {A.}~\bibnamefont
  {Kitaev}},\ }\href {\doibase 10.1103/PhysRevB.83.075103} {\bibfield
  {journal} {\bibinfo  {journal} {Phys. Rev. B}\ }\textbf {\bibinfo {volume}
  {83}},\ \bibinfo {pages} {075103} (\bibinfo {year} {2011})}\BibitemShut
  {NoStop}%
\bibitem [{\citenamefont {Chen}\ \emph {et~al.}(2012)\citenamefont {Chen},
  \citenamefont {Gu}, \citenamefont {Liu},\ and\ \citenamefont
  {Wen}}]{Chen1604}%
  \BibitemOpen
  \bibfield  {author} {\bibinfo {author} {\bibfnamefont {X.}~\bibnamefont
  {Chen}}, \bibinfo {author} {\bibfnamefont {Z.-C.}\ \bibnamefont {Gu}},
  \bibinfo {author} {\bibfnamefont {Z.-X.}\ \bibnamefont {Liu}}, \ and\
  \bibinfo {author} {\bibfnamefont {X.-G.}\ \bibnamefont {Wen}},\ }\href
  {\doibase 10.1126/science.1227224} {\bibfield  {journal} {\bibinfo  {journal}
  {Science}\ }\textbf {\bibinfo {volume} {338}},\ \bibinfo {pages} {1604}
  (\bibinfo {year} {2012})}\BibitemShut {NoStop}%
\bibitem [{\citenamefont {Kane}\ and\ \citenamefont
  {Mele}(2005)}]{PhysRevLett.95.226801}%
  \BibitemOpen
  \bibfield  {author} {\bibinfo {author} {\bibfnamefont {C.~L.}\ \bibnamefont
  {Kane}}\ and\ \bibinfo {author} {\bibfnamefont {E.~J.}\ \bibnamefont
  {Mele}},\ }\href {\doibase 10.1103/PhysRevLett.95.226801} {\bibfield
  {journal} {\bibinfo  {journal} {Phys. Rev. Lett.}\ }\textbf {\bibinfo
  {volume} {95}},\ \bibinfo {pages} {226801} (\bibinfo {year}
  {2005})}\BibitemShut {NoStop}%
\bibitem [{\citenamefont {Bernevig}\ \emph {et~al.}(2006)\citenamefont
  {Bernevig}, \citenamefont {Hughes},\ and\ \citenamefont
  {Zhang}}]{Bernevig1757}%
  \BibitemOpen
  \bibfield  {author} {\bibinfo {author} {\bibfnamefont {B.~A.}\ \bibnamefont
  {Bernevig}}, \bibinfo {author} {\bibfnamefont {T.~L.}\ \bibnamefont
  {Hughes}}, \ and\ \bibinfo {author} {\bibfnamefont {S.-C.}\ \bibnamefont
  {Zhang}},\ }\href {\doibase 10.1126/science.1133734} {\bibfield  {journal}
  {\bibinfo  {journal} {Science}\ }\textbf {\bibinfo {volume} {314}},\ \bibinfo
  {pages} {1757} (\bibinfo {year} {2006})}\BibitemShut {NoStop}%
\bibitem [{\citenamefont {Fu}\ \emph {et~al.}(2007)\citenamefont {Fu},
  \citenamefont {Kane},\ and\ \citenamefont {Mele}}]{PhysRevLett.98.106803}%
  \BibitemOpen
  \bibfield  {author} {\bibinfo {author} {\bibfnamefont {L.}~\bibnamefont
  {Fu}}, \bibinfo {author} {\bibfnamefont {C.~L.}\ \bibnamefont {Kane}}, \ and\
  \bibinfo {author} {\bibfnamefont {E.~J.}\ \bibnamefont {Mele}},\ }\href
  {\doibase 10.1103/PhysRevLett.98.106803} {\bibfield  {journal} {\bibinfo
  {journal} {Phys. Rev. Lett.}\ }\textbf {\bibinfo {volume} {98}},\ \bibinfo
  {pages} {106803} (\bibinfo {year} {2007})}\BibitemShut {NoStop}%
\bibitem [{\citenamefont {Moore}\ and\ \citenamefont
  {Balents}(2007)}]{PhysRevB.75.121306}%
  \BibitemOpen
  \bibfield  {author} {\bibinfo {author} {\bibfnamefont {J.~E.}\ \bibnamefont
  {Moore}}\ and\ \bibinfo {author} {\bibfnamefont {L.}~\bibnamefont
  {Balents}},\ }\href {\doibase 10.1103/PhysRevB.75.121306} {\bibfield
  {journal} {\bibinfo  {journal} {Phys. Rev. B}\ }\textbf {\bibinfo {volume}
  {75}},\ \bibinfo {pages} {121306} (\bibinfo {year} {2007})}\BibitemShut
  {NoStop}%
\bibitem [{\citenamefont {Roy}(2009)}]{PhysRevB.79.195322}%
  \BibitemOpen
  \bibfield  {author} {\bibinfo {author} {\bibfnamefont {R.}~\bibnamefont
  {Roy}},\ }\href {\doibase 10.1103/PhysRevB.79.195322} {\bibfield  {journal}
  {\bibinfo  {journal} {Phys. Rev. B}\ }\textbf {\bibinfo {volume} {79}},\
  \bibinfo {pages} {195322} (\bibinfo {year} {2009})}\BibitemShut {NoStop}%
\bibitem [{\citenamefont {Hasan}\ and\ \citenamefont
  {Kane}(2010)}]{RevModPhys.82.3045}%
  \BibitemOpen
  \bibfield  {author} {\bibinfo {author} {\bibfnamefont {M.~Z.}\ \bibnamefont
  {Hasan}}\ and\ \bibinfo {author} {\bibfnamefont {C.~L.}\ \bibnamefont
  {Kane}},\ }\href {\doibase 10.1103/RevModPhys.82.3045} {\bibfield  {journal}
  {\bibinfo  {journal} {Rev. Mod. Phys.}\ }\textbf {\bibinfo {volume} {82}},\
  \bibinfo {pages} {3045} (\bibinfo {year} {2010})}\BibitemShut {NoStop}%
\bibitem [{\citenamefont {Wen}(2002)}]{PhysRevB.65.165113}%
  \BibitemOpen
  \bibfield  {author} {\bibinfo {author} {\bibfnamefont {X.-G.}\ \bibnamefont
  {Wen}},\ }\href {\doibase 10.1103/PhysRevB.65.165113} {\bibfield  {journal}
  {\bibinfo  {journal} {Phys. Rev. B}\ }\textbf {\bibinfo {volume} {65}},\
  \bibinfo {pages} {165113} (\bibinfo {year} {2002})}\BibitemShut {NoStop}%
\bibitem [{\citenamefont {Maciejko}\ \emph {et~al.}(2010)\citenamefont
  {Maciejko}, \citenamefont {Qi}, \citenamefont {Karch},\ and\ \citenamefont
  {Zhang}}]{PhysRevLett.105.246809}%
  \BibitemOpen
  \bibfield  {author} {\bibinfo {author} {\bibfnamefont {J.}~\bibnamefont
  {Maciejko}}, \bibinfo {author} {\bibfnamefont {X.-L.}\ \bibnamefont {Qi}},
  \bibinfo {author} {\bibfnamefont {A.}~\bibnamefont {Karch}}, \ and\ \bibinfo
  {author} {\bibfnamefont {S.-C.}\ \bibnamefont {Zhang}},\ }\href {\doibase
  10.1103/PhysRevLett.105.246809} {\bibfield  {journal} {\bibinfo  {journal}
  {Phys. Rev. Lett.}\ }\textbf {\bibinfo {volume} {105}},\ \bibinfo {pages}
  {246809} (\bibinfo {year} {2010})}\BibitemShut {NoStop}%
\bibitem [{\citenamefont {Swingle}\ \emph {et~al.}(2011)\citenamefont
  {Swingle}, \citenamefont {Barkeshli}, \citenamefont {McGreevy},\ and\
  \citenamefont {Senthil}}]{PhysRevB.83.195139}%
  \BibitemOpen
  \bibfield  {author} {\bibinfo {author} {\bibfnamefont {B.}~\bibnamefont
  {Swingle}}, \bibinfo {author} {\bibfnamefont {M.}~\bibnamefont {Barkeshli}},
  \bibinfo {author} {\bibfnamefont {J.}~\bibnamefont {McGreevy}}, \ and\
  \bibinfo {author} {\bibfnamefont {T.}~\bibnamefont {Senthil}},\ }\href
  {\doibase 10.1103/PhysRevB.83.195139} {\bibfield  {journal} {\bibinfo
  {journal} {Phys. Rev. B}\ }\textbf {\bibinfo {volume} {83}},\ \bibinfo
  {pages} {195139} (\bibinfo {year} {2011})}\BibitemShut {NoStop}%
\bibitem [{\citenamefont {Levin}\ and\ \citenamefont
  {Stern}(2012)}]{PhysRevB.86.115131}%
  \BibitemOpen
  \bibfield  {author} {\bibinfo {author} {\bibfnamefont {M.}~\bibnamefont
  {Levin}}\ and\ \bibinfo {author} {\bibfnamefont {A.}~\bibnamefont {Stern}},\
  }\href {\doibase 10.1103/PhysRevB.86.115131} {\bibfield  {journal} {\bibinfo
  {journal} {Phys. Rev. B}\ }\textbf {\bibinfo {volume} {86}},\ \bibinfo
  {pages} {115131} (\bibinfo {year} {2012})}\BibitemShut {NoStop}%
\bibitem [{\citenamefont {Mesaros}\ and\ \citenamefont
  {Ran}(2013)}]{PhysRevB.87.155115}%
  \BibitemOpen
  \bibfield  {author} {\bibinfo {author} {\bibfnamefont {A.}~\bibnamefont
  {Mesaros}}\ and\ \bibinfo {author} {\bibfnamefont {Y.}~\bibnamefont {Ran}},\
  }\href {\doibase 10.1103/PhysRevB.87.155115} {\bibfield  {journal} {\bibinfo
  {journal} {Phys. Rev. B}\ }\textbf {\bibinfo {volume} {87}},\ \bibinfo
  {pages} {155115} (\bibinfo {year} {2013})}\BibitemShut {NoStop}%
\bibitem [{\citenamefont {Fleishman}\ and\ \citenamefont
  {Anderson}(1980)}]{FleishmanAnderson}%
  \BibitemOpen
  \bibfield  {author} {\bibinfo {author} {\bibfnamefont {L.}~\bibnamefont
  {Fleishman}}\ and\ \bibinfo {author} {\bibfnamefont {P.}~\bibnamefont
  {Anderson}},\ }\href {\doibase 10.1103/PhysRevB.21.2366} {\bibfield
  {journal} {\bibinfo  {journal} {Phys. Rev. B}\ }\textbf {\bibinfo {volume}
  {21}},\ \bibinfo {pages} {2366} (\bibinfo {year} {1980})}\BibitemShut
  {NoStop}%
\bibitem [{\citenamefont {Gornyi}\ \emph {et~al.}(2005)\citenamefont {Gornyi},
  \citenamefont {Mirlin},\ and\ \citenamefont {Polyakov}}]{Gornyi}%
  \BibitemOpen
  \bibfield  {author} {\bibinfo {author} {\bibfnamefont {I.}~\bibnamefont
  {Gornyi}}, \bibinfo {author} {\bibfnamefont {A.}~\bibnamefont {Mirlin}}, \
  and\ \bibinfo {author} {\bibfnamefont {D.}~\bibnamefont {Polyakov}},\ }\href
  {\doibase 10.1103/PhysRevLett.95.206603} {\bibfield  {journal} {\bibinfo
  {journal} {Phys. Rev. Lett.}\ }\textbf {\bibinfo {volume} {95}},\ \bibinfo
  {pages} {206603} (\bibinfo {year} {2005})}\BibitemShut {NoStop}%
\bibitem [{\citenamefont {Basko}\ \emph {et~al.}(2006)\citenamefont {Basko},
  \citenamefont {Aleiner},\ and\ \citenamefont {Altshuler}}]{BAA}%
  \BibitemOpen
  \bibfield  {author} {\bibinfo {author} {\bibfnamefont {D.}~\bibnamefont
  {Basko}}, \bibinfo {author} {\bibfnamefont {I.}~\bibnamefont {Aleiner}}, \
  and\ \bibinfo {author} {\bibfnamefont {B.}~\bibnamefont {Altshuler}},\ }\href
  {\doibase http://dx.doi.org/10.1016/j.aop.2005.11.014} {\bibfield  {journal}
  {\bibinfo  {journal} {Annals of Physics}\ }\textbf {\bibinfo {volume}
  {321}},\ \bibinfo {pages} {1126 } (\bibinfo {year} {2006})}\BibitemShut
  {NoStop}%
\bibitem [{\citenamefont {Oganesyan}\ and\ \citenamefont
  {Huse}(2007)}]{PhysRevB.75.155111}%
  \BibitemOpen
  \bibfield  {author} {\bibinfo {author} {\bibfnamefont {V.}~\bibnamefont
  {Oganesyan}}\ and\ \bibinfo {author} {\bibfnamefont {D.~A.}\ \bibnamefont
  {Huse}},\ }\href {\doibase 10.1103/PhysRevB.75.155111} {\bibfield  {journal}
  {\bibinfo  {journal} {Phys. Rev. B}\ }\textbf {\bibinfo {volume} {75}},\
  \bibinfo {pages} {155111} (\bibinfo {year} {2007})}\BibitemShut {NoStop}%
\bibitem [{\citenamefont {Pal}\ and\ \citenamefont {Huse}(2010)}]{PalHuse}%
  \BibitemOpen
  \bibfield  {author} {\bibinfo {author} {\bibfnamefont {A.}~\bibnamefont
  {Pal}}\ and\ \bibinfo {author} {\bibfnamefont {D.~A.}\ \bibnamefont {Huse}},\
  }\href {\doibase 10.1103/PhysRevB.82.174411} {\bibfield  {journal} {\bibinfo
  {journal} {Phys. Rev. B}\ }\textbf {\bibinfo {volume} {82}},\ \bibinfo
  {pages} {174411} (\bibinfo {year} {2010})}\BibitemShut {NoStop}%
\bibitem [{\citenamefont {Bauer}\ and\ \citenamefont
  {Nayak}(2013)}]{BauerNayak}%
  \BibitemOpen
  \bibfield  {author} {\bibinfo {author} {\bibfnamefont {B.}~\bibnamefont
  {Bauer}}\ and\ \bibinfo {author} {\bibfnamefont {C.}~\bibnamefont {Nayak}},\
  }\href {http://stacks.iop.org/1742-5468/2013/i=09/a=P09005} {\bibfield
  {journal} {\bibinfo  {journal} {Journal of Statistical Mechanics}\ }\textbf
  {\bibinfo {volume} {2013}},\ \bibinfo {pages} {P09005} (\bibinfo {year}
  {2013})}\BibitemShut {NoStop}%
\bibitem [{\citenamefont {Bardarson}\ \emph {et~al.}(2012)\citenamefont
  {Bardarson}, \citenamefont {Pollmann},\ and\ \citenamefont
  {Moore}}]{PhysRevLett.109.017202}%
  \BibitemOpen
  \bibfield  {author} {\bibinfo {author} {\bibfnamefont {J.~H.}\ \bibnamefont
  {Bardarson}}, \bibinfo {author} {\bibfnamefont {F.}~\bibnamefont {Pollmann}},
  \ and\ \bibinfo {author} {\bibfnamefont {J.~E.}\ \bibnamefont {Moore}},\
  }\href {\doibase 10.1103/PhysRevLett.109.017202} {\bibfield  {journal}
  {\bibinfo  {journal} {Phys. Rev. Lett.}\ }\textbf {\bibinfo {volume} {109}},\
  \bibinfo {pages} {017202} (\bibinfo {year} {2012})}\BibitemShut {NoStop}%
\bibitem [{\citenamefont {Serbyn}\ \emph
  {et~al.}(2013{\natexlab{a}})\citenamefont {Serbyn}, \citenamefont
  {Papi\ifmmode~\acute{c}\else \'{c}\fi{}},\ and\ \citenamefont
  {Abanin}}]{PhysRevLett.110.260601}%
  \BibitemOpen
  \bibfield  {author} {\bibinfo {author} {\bibfnamefont {M.}~\bibnamefont
  {Serbyn}}, \bibinfo {author} {\bibfnamefont {Z.}~\bibnamefont
  {Papi\ifmmode~\acute{c}\else \'{c}\fi{}}}, \ and\ \bibinfo {author}
  {\bibfnamefont {D.~A.}\ \bibnamefont {Abanin}},\ }\href {\doibase
  10.1103/PhysRevLett.110.260601} {\bibfield  {journal} {\bibinfo  {journal}
  {Phys. Rev. Lett.}\ }\textbf {\bibinfo {volume} {110}},\ \bibinfo {pages}
  {260601} (\bibinfo {year} {2013}{\natexlab{a}})}\BibitemShut {NoStop}%
\bibitem [{\citenamefont {Serbyn}\ \emph
  {et~al.}(2013{\natexlab{b}})\citenamefont {Serbyn}, \citenamefont
  {Papi\ifmmode~\acute{c}\else \'{c}\fi{}},\ and\ \citenamefont
  {Abanin}}]{PhysRevLett.111.127201}%
  \BibitemOpen
  \bibfield  {author} {\bibinfo {author} {\bibfnamefont {M.}~\bibnamefont
  {Serbyn}}, \bibinfo {author} {\bibfnamefont {Z.}~\bibnamefont
  {Papi\ifmmode~\acute{c}\else \'{c}\fi{}}}, \ and\ \bibinfo {author}
  {\bibfnamefont {D.~A.}\ \bibnamefont {Abanin}},\ }\href {\doibase
  10.1103/PhysRevLett.111.127201} {\bibfield  {journal} {\bibinfo  {journal}
  {Phys. Rev. Lett.}\ }\textbf {\bibinfo {volume} {111}},\ \bibinfo {pages}
  {127201} (\bibinfo {year} {2013}{\natexlab{b}})}\BibitemShut {NoStop}%
\bibitem [{\citenamefont {Huse}\ \emph {et~al.}(2014)\citenamefont {Huse},
  \citenamefont {Nandkishore},\ and\ \citenamefont
  {Oganesyan}}]{PhysRevB.90.174202}%
  \BibitemOpen
  \bibfield  {author} {\bibinfo {author} {\bibfnamefont {D.~A.}\ \bibnamefont
  {Huse}}, \bibinfo {author} {\bibfnamefont {R.}~\bibnamefont {Nandkishore}}, \
  and\ \bibinfo {author} {\bibfnamefont {V.}~\bibnamefont {Oganesyan}},\ }\href
  {\doibase 10.1103/PhysRevB.90.174202} {\bibfield  {journal} {\bibinfo
  {journal} {Phys. Rev. B}\ }\textbf {\bibinfo {volume} {90}},\ \bibinfo
  {pages} {174202} (\bibinfo {year} {2014})}\BibitemShut {NoStop}%
\bibitem [{\citenamefont {Bahri}\ \emph {et~al.}(2015)\citenamefont {Bahri},
  \citenamefont {Vosk}, \citenamefont {Altman},\ and\ \citenamefont
  {Vishwanath}}]{BahriMBLSPT}%
  \BibitemOpen
  \bibfield  {author} {\bibinfo {author} {\bibfnamefont {Y.}~\bibnamefont
  {Bahri}}, \bibinfo {author} {\bibfnamefont {R.}~\bibnamefont {Vosk}},
  \bibinfo {author} {\bibfnamefont {E.}~\bibnamefont {Altman}}, \ and\ \bibinfo
  {author} {\bibfnamefont {A.}~\bibnamefont {Vishwanath}},\ }\href
  {http://dx.doi.org/10.1038/ncomms8341} {\bibfield  {journal} {\bibinfo
  {journal} {Nat Commun}\ }\textbf {\bibinfo {volume} {6}} (\bibinfo {year}
  {2015})}\BibitemShut {NoStop}%
\bibitem [{\citenamefont {Serbyn}\ \emph {et~al.}(2014)\citenamefont {Serbyn},
  \citenamefont {Knap}, \citenamefont {Gopalakrishnan}, \citenamefont
  {Papi\ifmmode~\acute{c}\else \'{c}\fi{}}, \citenamefont {Yao}, \citenamefont
  {Laumann}, \citenamefont {Abanin}, \citenamefont {Lukin},\ and\ \citenamefont
  {Demler}}]{PhysRevLett.113.147204}%
  \BibitemOpen
  \bibfield  {author} {\bibinfo {author} {\bibfnamefont {M.}~\bibnamefont
  {Serbyn}}, \bibinfo {author} {\bibfnamefont {M.}~\bibnamefont {Knap}},
  \bibinfo {author} {\bibfnamefont {S.}~\bibnamefont {Gopalakrishnan}},
  \bibinfo {author} {\bibfnamefont {Z.}~\bibnamefont
  {Papi\ifmmode~\acute{c}\else \'{c}\fi{}}}, \bibinfo {author} {\bibfnamefont
  {N.~Y.}\ \bibnamefont {Yao}}, \bibinfo {author} {\bibfnamefont {C.~R.}\
  \bibnamefont {Laumann}}, \bibinfo {author} {\bibfnamefont {D.~A.}\
  \bibnamefont {Abanin}}, \bibinfo {author} {\bibfnamefont {M.~D.}\
  \bibnamefont {Lukin}}, \ and\ \bibinfo {author} {\bibfnamefont {E.~A.}\
  \bibnamefont {Demler}},\ }\href {\doibase 10.1103/PhysRevLett.113.147204}
  {\bibfield  {journal} {\bibinfo  {journal} {Phys. Rev. Lett.}\ }\textbf
  {\bibinfo {volume} {113}},\ \bibinfo {pages} {147204} (\bibinfo {year}
  {2014})}\BibitemShut {NoStop}%
\bibitem [{\citenamefont {Vasseur}\ \emph
  {et~al.}(2015{\natexlab{a}})\citenamefont {Vasseur}, \citenamefont
  {Parameswaran},\ and\ \citenamefont {Moore}}]{PhysRevB.91.140202}%
  \BibitemOpen
  \bibfield  {author} {\bibinfo {author} {\bibfnamefont {R.}~\bibnamefont
  {Vasseur}}, \bibinfo {author} {\bibfnamefont {S.~A.}\ \bibnamefont
  {Parameswaran}}, \ and\ \bibinfo {author} {\bibfnamefont {J.~E.}\
  \bibnamefont {Moore}},\ }\href {\doibase 10.1103/PhysRevB.91.140202}
  {\bibfield  {journal} {\bibinfo  {journal} {Phys. Rev. B}\ }\textbf {\bibinfo
  {volume} {91}},\ \bibinfo {pages} {140202} (\bibinfo {year}
  {2015}{\natexlab{a}})}\BibitemShut {NoStop}%
\bibitem [{\citenamefont {Huse}\ \emph {et~al.}(2013)\citenamefont {Huse},
  \citenamefont {Nandkishore}, \citenamefont {Oganesyan}, \citenamefont {Pal},\
  and\ \citenamefont {Sondhi}}]{HuseMBLQuantumOrder}%
  \BibitemOpen
  \bibfield  {author} {\bibinfo {author} {\bibfnamefont {D.~A.}\ \bibnamefont
  {Huse}}, \bibinfo {author} {\bibfnamefont {R.}~\bibnamefont {Nandkishore}},
  \bibinfo {author} {\bibfnamefont {V.}~\bibnamefont {Oganesyan}}, \bibinfo
  {author} {\bibfnamefont {A.}~\bibnamefont {Pal}}, \ and\ \bibinfo {author}
  {\bibfnamefont {S.~L.}\ \bibnamefont {Sondhi}},\ }\href {\doibase
  10.1103/PhysRevB.88.014206} {\bibfield  {journal} {\bibinfo  {journal} {Phys.
  Rev. B}\ }\textbf {\bibinfo {volume} {88}},\ \bibinfo {pages} {014206}
  (\bibinfo {year} {2013})}\BibitemShut {NoStop}%
\bibitem [{\citenamefont {Pekker}\ \emph {et~al.}(2014)\citenamefont {Pekker},
  \citenamefont {Refael}, \citenamefont {Altman}, \citenamefont {Demler},\ and\
  \citenamefont {Oganesyan}}]{PekkerRSRGX}%
  \BibitemOpen
  \bibfield  {author} {\bibinfo {author} {\bibfnamefont {D.}~\bibnamefont
  {Pekker}}, \bibinfo {author} {\bibfnamefont {G.}~\bibnamefont {Refael}},
  \bibinfo {author} {\bibfnamefont {E.}~\bibnamefont {Altman}}, \bibinfo
  {author} {\bibfnamefont {E.}~\bibnamefont {Demler}}, \ and\ \bibinfo {author}
  {\bibfnamefont {V.}~\bibnamefont {Oganesyan}},\ }\href {\doibase
  10.1103/PhysRevX.4.011052} {\bibfield  {journal} {\bibinfo  {journal} {Phys.
  Rev. X}\ }\textbf {\bibinfo {volume} {4}},\ \bibinfo {pages} {011052}
  (\bibinfo {year} {2014})}\BibitemShut {NoStop}%
\bibitem [{\citenamefont {Chandran}\ \emph {et~al.}(2014)\citenamefont
  {Chandran}, \citenamefont {Khemani}, \citenamefont {Laumann},\ and\
  \citenamefont {Sondhi}}]{PhysRevB.89.144201}%
  \BibitemOpen
  \bibfield  {author} {\bibinfo {author} {\bibfnamefont {A.}~\bibnamefont
  {Chandran}}, \bibinfo {author} {\bibfnamefont {V.}~\bibnamefont {Khemani}},
  \bibinfo {author} {\bibfnamefont {C.~R.}\ \bibnamefont {Laumann}}, \ and\
  \bibinfo {author} {\bibfnamefont {S.~L.}\ \bibnamefont {Sondhi}},\ }\href
  {\doibase 10.1103/PhysRevB.89.144201} {\bibfield  {journal} {\bibinfo
  {journal} {Phys. Rev. B}\ }\textbf {\bibinfo {volume} {89}},\ \bibinfo
  {pages} {144201} (\bibinfo {year} {2014})}\BibitemShut {NoStop}%
\bibitem [{\citenamefont {{Potter}}\ and\ \citenamefont
  {{Vishwanath}}(2015)}]{2015arXiv150600592P}%
  \BibitemOpen
  \bibfield  {author} {\bibinfo {author} {\bibfnamefont {A.~C.}\ \bibnamefont
  {{Potter}}}\ and\ \bibinfo {author} {\bibfnamefont {A.}~\bibnamefont
  {{Vishwanath}}},\ }\href@noop {} {\bibfield  {journal} {\bibinfo  {journal}
  {ArXiv e-prints}\ } (\bibinfo {year} {2015})},\ \Eprint
  {http://arxiv.org/abs/1506.00592} {arXiv:1506.00592 [cond-mat.dis-nn]}
  \BibitemShut {NoStop}%
\bibitem [{\citenamefont {{Slagle}}\ \emph {et~al.}(2015)\citenamefont
  {{Slagle}}, \citenamefont {{Bi}}, \citenamefont {{You}},\ and\ \citenamefont
  {{Xu}}}]{2015arXiv150505147S}%
  \BibitemOpen
  \bibfield  {author} {\bibinfo {author} {\bibfnamefont {K.}~\bibnamefont
  {{Slagle}}}, \bibinfo {author} {\bibfnamefont {Z.}~\bibnamefont {{Bi}}},
  \bibinfo {author} {\bibfnamefont {Y.-Z.}\ \bibnamefont {{You}}}, \ and\
  \bibinfo {author} {\bibfnamefont {C.}~\bibnamefont {{Xu}}},\ }\href@noop {}
  {\bibfield  {journal} {\bibinfo  {journal} {ArXiv e-prints}\ } (\bibinfo
  {year} {2015})},\ \Eprint {http://arxiv.org/abs/1505.05147} {arXiv:1505.05147
  [cond-mat.str-el]} \BibitemShut {NoStop}%
\bibitem [{\citenamefont {Vasseur}\ \emph
  {et~al.}(2015{\natexlab{b}})\citenamefont {Vasseur}, \citenamefont {Potter},\
  and\ \citenamefont {Parameswaran}}]{QCGPRL}%
  \BibitemOpen
  \bibfield  {author} {\bibinfo {author} {\bibfnamefont {R.}~\bibnamefont
  {Vasseur}}, \bibinfo {author} {\bibfnamefont {A.~C.}\ \bibnamefont {Potter}},
  \ and\ \bibinfo {author} {\bibfnamefont {S.~A.}\ \bibnamefont
  {Parameswaran}},\ }\href {\doibase 10.1103/PhysRevLett.114.217201} {\bibfield
   {journal} {\bibinfo  {journal} {Phys. Rev. Lett.}\ }\textbf {\bibinfo
  {volume} {114}},\ \bibinfo {pages} {217201} (\bibinfo {year}
  {2015}{\natexlab{b}})}\BibitemShut {NoStop}%
\bibitem [{\citenamefont {Vasseur}\ \emph {et~al.}(2016)\citenamefont
  {Vasseur}, \citenamefont {Friedman}, \citenamefont {Parameswaran},\ and\
  \citenamefont {Potter}}]{2015arXiv151004282V}%
  \BibitemOpen
  \bibfield  {author} {\bibinfo {author} {\bibfnamefont {R.}~\bibnamefont
  {Vasseur}}, \bibinfo {author} {\bibfnamefont {A.~J.}\ \bibnamefont
  {Friedman}}, \bibinfo {author} {\bibfnamefont {S.~A.}\ \bibnamefont
  {Parameswaran}}, \ and\ \bibinfo {author} {\bibfnamefont {A.~C.}\
  \bibnamefont {Potter}},\ }\href {\doibase 10.1103/PhysRevB.93.134207}
  {\bibfield  {journal} {\bibinfo  {journal} {Phys. Rev. B}\ }\textbf {\bibinfo
  {volume} {93}},\ \bibinfo {pages} {134207} (\bibinfo {year}
  {2016})}\BibitemShut {NoStop}%
\bibitem [{\citenamefont {{Potter}}\ \emph {et~al.}(2016)\citenamefont
  {{Potter}}, \citenamefont {{Morimoto}},\ and\ \citenamefont
  {{Vishwanath}}}]{2016arXiv160205194P}%
  \BibitemOpen
  \bibfield  {author} {\bibinfo {author} {\bibfnamefont {A.~C.}\ \bibnamefont
  {{Potter}}}, \bibinfo {author} {\bibfnamefont {T.}~\bibnamefont
  {{Morimoto}}}, \ and\ \bibinfo {author} {\bibfnamefont {A.}~\bibnamefont
  {{Vishwanath}}},\ }\href@noop {} {\bibfield  {journal} {\bibinfo  {journal}
  {ArXiv e-prints}\ } (\bibinfo {year} {2016})},\ \Eprint
  {http://arxiv.org/abs/1602.05194} {arXiv:1602.05194 [cond-mat.str-el]}
  \BibitemShut {NoStop}%
\bibitem [{\citenamefont {Vosk}\ and\ \citenamefont
  {Altman}(2013)}]{VoskAltmanPRL13}%
  \BibitemOpen
  \bibfield  {author} {\bibinfo {author} {\bibfnamefont {R.}~\bibnamefont
  {Vosk}}\ and\ \bibinfo {author} {\bibfnamefont {E.}~\bibnamefont {Altman}},\
  }\href {\doibase 10.1103/PhysRevLett.110.067204} {\bibfield  {journal}
  {\bibinfo  {journal} {Phys. Rev. Lett.}\ }\textbf {\bibinfo {volume} {110}},\
  \bibinfo {pages} {067204} (\bibinfo {year} {2013})}\BibitemShut {NoStop}%
\bibitem [{\citenamefont {{Imbrie}}(2014)}]{2014arXiv1403.7837I}%
  \BibitemOpen
  \bibfield  {author} {\bibinfo {author} {\bibfnamefont {J.~Z.}\ \bibnamefont
  {{Imbrie}}},\ }\href@noop {} {\bibfield  {journal} {\bibinfo  {journal}
  {ArXiv e-prints}\ } (\bibinfo {year} {2014})},\ \Eprint
  {http://arxiv.org/abs/1403.7837} {arXiv:1403.7837 [math-ph]} \BibitemShut
  {NoStop}%
\bibitem [{\citenamefont {Znidaric}\ \emph {et~al.}(2008)\citenamefont
  {Znidaric}, \citenamefont {Prosen},\ and\ \citenamefont
  {Prelovsek}}]{PhysRevB.77.064426}%
  \BibitemOpen
  \bibfield  {author} {\bibinfo {author} {\bibfnamefont {M.}~\bibnamefont
  {Znidaric}}, \bibinfo {author} {\bibfnamefont {T.}~\bibnamefont {Prosen}}, \
  and\ \bibinfo {author} {\bibfnamefont {P.}~\bibnamefont {Prelovsek}},\ }\href
  {\doibase 10.1103/PhysRevB.77.064426} {\bibfield  {journal} {\bibinfo
  {journal} {Phys. Rev. B}\ }\textbf {\bibinfo {volume} {77}},\ \bibinfo
  {pages} {064426} (\bibinfo {year} {2008})}\BibitemShut {NoStop}%
\bibitem [{\citenamefont {Lazarides}\ \emph {et~al.}(2015)\citenamefont
  {Lazarides}, \citenamefont {Das},\ and\ \citenamefont
  {Moessner}}]{PhysRevLett.115.030402}%
  \BibitemOpen
  \bibfield  {author} {\bibinfo {author} {\bibfnamefont {A.}~\bibnamefont
  {Lazarides}}, \bibinfo {author} {\bibfnamefont {A.}~\bibnamefont {Das}}, \
  and\ \bibinfo {author} {\bibfnamefont {R.}~\bibnamefont {Moessner}},\ }\href
  {\doibase 10.1103/PhysRevLett.115.030402} {\bibfield  {journal} {\bibinfo
  {journal} {Phys. Rev. Lett.}\ }\textbf {\bibinfo {volume} {115}},\ \bibinfo
  {pages} {030402} (\bibinfo {year} {2015})}\BibitemShut {NoStop}%
\bibitem [{\citenamefont {Ponte}\ \emph {et~al.}(2015)\citenamefont {Ponte},
  \citenamefont {Papi\ifmmode~\acute{c}\else \'{c}\fi{}}, \citenamefont
  {Huveneers},\ and\ \citenamefont {Abanin}}]{PhysRevLett.114.140401}%
  \BibitemOpen
  \bibfield  {author} {\bibinfo {author} {\bibfnamefont {P.}~\bibnamefont
  {Ponte}}, \bibinfo {author} {\bibfnamefont {Z.}~\bibnamefont
  {Papi\ifmmode~\acute{c}\else \'{c}\fi{}}}, \bibinfo {author} {\bibfnamefont
  {F.~m.~c.}\ \bibnamefont {Huveneers}}, \ and\ \bibinfo {author}
  {\bibfnamefont {D.~A.}\ \bibnamefont {Abanin}},\ }\href {\doibase
  10.1103/PhysRevLett.114.140401} {\bibfield  {journal} {\bibinfo  {journal}
  {Phys. Rev. Lett.}\ }\textbf {\bibinfo {volume} {114}},\ \bibinfo {pages}
  {140401} (\bibinfo {year} {2015})}\BibitemShut {NoStop}%
\bibitem [{\citenamefont {De~Roeck}\ \emph {et~al.}(2016)\citenamefont
  {De~Roeck}, \citenamefont {Huveneers}, \citenamefont {M\"uller},\ and\
  \citenamefont {Schiulaz}}]{PhysRevB.93.014203}%
  \BibitemOpen
  \bibfield  {author} {\bibinfo {author} {\bibfnamefont {W.}~\bibnamefont
  {De~Roeck}}, \bibinfo {author} {\bibfnamefont {F.}~\bibnamefont {Huveneers}},
  \bibinfo {author} {\bibfnamefont {M.}~\bibnamefont {M\"uller}}, \ and\
  \bibinfo {author} {\bibfnamefont {M.}~\bibnamefont {Schiulaz}},\ }\href
  {\doibase 10.1103/PhysRevB.93.014203} {\bibfield  {journal} {\bibinfo
  {journal} {Phys. Rev. B}\ }\textbf {\bibinfo {volume} {93}},\ \bibinfo
  {pages} {014203} (\bibinfo {year} {2016})}\BibitemShut {NoStop}%
\bibitem [{\citenamefont {{Geraedts}}\ \emph {et~al.}(2016)\citenamefont
  {{Geraedts}}, \citenamefont {{Bhatt}},\ and\ \citenamefont
  {{Nandkishore}}}]{2016arXiv160801328G}%
  \BibitemOpen
  \bibfield  {author} {\bibinfo {author} {\bibfnamefont {S.~D.}\ \bibnamefont
  {{Geraedts}}}, \bibinfo {author} {\bibfnamefont {R.~N.}\ \bibnamefont
  {{Bhatt}}}, \ and\ \bibinfo {author} {\bibfnamefont {R.}~\bibnamefont
  {{Nandkishore}}},\ }\href@noop {} {\bibfield  {journal} {\bibinfo  {journal}
  {ArXiv e-prints}\ } (\bibinfo {year} {2016})},\ \Eprint
  {http://arxiv.org/abs/1608.01328} {arXiv:1608.01328 [cond-mat.stat-mech]}
  \BibitemShut {NoStop}%
\bibitem [{\citenamefont {Chen}\ \emph {et~al.}(2013)\citenamefont {Chen},
  \citenamefont {Gu}, \citenamefont {Liu},\ and\ \citenamefont
  {Wen}}]{PhysRevB.87.155114}%
  \BibitemOpen
  \bibfield  {author} {\bibinfo {author} {\bibfnamefont {X.}~\bibnamefont
  {Chen}}, \bibinfo {author} {\bibfnamefont {Z.-C.}\ \bibnamefont {Gu}},
  \bibinfo {author} {\bibfnamefont {Z.-X.}\ \bibnamefont {Liu}}, \ and\
  \bibinfo {author} {\bibfnamefont {X.-G.}\ \bibnamefont {Wen}},\ }\href
  {\doibase 10.1103/PhysRevB.87.155114} {\bibfield  {journal} {\bibinfo
  {journal} {Phys. Rev. B}\ }\textbf {\bibinfo {volume} {87}},\ \bibinfo
  {pages} {155114} (\bibinfo {year} {2013})}\BibitemShut {NoStop}%
\bibitem [{\citenamefont {Vosk}\ and\ \citenamefont
  {Altman}(2014)}]{PhysRevLett.112.217204}%
  \BibitemOpen
  \bibfield  {author} {\bibinfo {author} {\bibfnamefont {R.}~\bibnamefont
  {Vosk}}\ and\ \bibinfo {author} {\bibfnamefont {E.}~\bibnamefont {Altman}},\
  }\href {\doibase 10.1103/PhysRevLett.112.217204} {\bibfield  {journal}
  {\bibinfo  {journal} {Phys. Rev. Lett.}\ }\textbf {\bibinfo {volume} {112}},\
  \bibinfo {pages} {217204} (\bibinfo {year} {2014})}\BibitemShut {NoStop}%
\bibitem [{\citenamefont {Nandkishore}\ and\ \citenamefont
  {Potter}(2014)}]{NandkishorePotterScaling}%
  \BibitemOpen
  \bibfield  {author} {\bibinfo {author} {\bibfnamefont {R.}~\bibnamefont
  {Nandkishore}}\ and\ \bibinfo {author} {\bibfnamefont {A.~C.}\ \bibnamefont
  {Potter}},\ }\href {\doibase 10.1103/PhysRevB.90.195115} {\bibfield
  {journal} {\bibinfo  {journal} {Phys. Rev. B}\ }\textbf {\bibinfo {volume}
  {90}},\ \bibinfo {pages} {195115} (\bibinfo {year} {2014})}\BibitemShut
  {NoStop}%
\bibitem [{\citenamefont {You}\ \emph {et~al.}(2016)\citenamefont {You},
  \citenamefont {Qi},\ and\ \citenamefont {Xu}}]{PhysRevB.93.104205}%
  \BibitemOpen
  \bibfield  {author} {\bibinfo {author} {\bibfnamefont {Y.-Z.}\ \bibnamefont
  {You}}, \bibinfo {author} {\bibfnamefont {X.-L.}\ \bibnamefont {Qi}}, \ and\
  \bibinfo {author} {\bibfnamefont {C.}~\bibnamefont {Xu}},\ }\href {\doibase
  10.1103/PhysRevB.93.104205} {\bibfield  {journal} {\bibinfo  {journal} {Phys.
  Rev. B}\ }\textbf {\bibinfo {volume} {93}},\ \bibinfo {pages} {104205}
  (\bibinfo {year} {2016})}\BibitemShut {NoStop}%
\bibitem [{\citenamefont {Oganesyan}\ \emph {et~al.}(2009)\citenamefont
  {Oganesyan}, \citenamefont {Pal},\ and\ \citenamefont
  {Huse}}]{PhysRevB.80.115104}%
  \BibitemOpen
  \bibfield  {author} {\bibinfo {author} {\bibfnamefont {V.}~\bibnamefont
  {Oganesyan}}, \bibinfo {author} {\bibfnamefont {A.}~\bibnamefont {Pal}}, \
  and\ \bibinfo {author} {\bibfnamefont {D.~A.}\ \bibnamefont {Huse}},\ }\href
  {\doibase 10.1103/PhysRevB.80.115104} {\bibfield  {journal} {\bibinfo
  {journal} {Phys. Rev. B}\ }\textbf {\bibinfo {volume} {80}},\ \bibinfo
  {pages} {115104} (\bibinfo {year} {2009})}\BibitemShut {NoStop}%
\bibitem [{\citenamefont {John}\ \emph {et~al.}(1983)\citenamefont {John},
  \citenamefont {Sompolinsky},\ and\ \citenamefont
  {Stephen}}]{PhysRevB.27.5592}%
  \BibitemOpen
  \bibfield  {author} {\bibinfo {author} {\bibfnamefont {S.}~\bibnamefont
  {John}}, \bibinfo {author} {\bibfnamefont {H.}~\bibnamefont {Sompolinsky}}, \
  and\ \bibinfo {author} {\bibfnamefont {M.~J.}\ \bibnamefont {Stephen}},\
  }\href {\doibase 10.1103/PhysRevB.27.5592} {\bibfield  {journal} {\bibinfo
  {journal} {Phys. Rev. B}\ }\textbf {\bibinfo {volume} {27}},\ \bibinfo
  {pages} {5592} (\bibinfo {year} {1983})}\BibitemShut {NoStop}%
\bibitem [{\citenamefont {Gurarie}\ and\ \citenamefont
  {Chalker}(2003)}]{PhysRevB.68.134207}%
  \BibitemOpen
  \bibfield  {author} {\bibinfo {author} {\bibfnamefont {V.}~\bibnamefont
  {Gurarie}}\ and\ \bibinfo {author} {\bibfnamefont {J.~T.}\ \bibnamefont
  {Chalker}},\ }\href {\doibase 10.1103/PhysRevB.68.134207} {\bibfield
  {journal} {\bibinfo  {journal} {Phys. Rev. B}\ }\textbf {\bibinfo {volume}
  {68}},\ \bibinfo {pages} {134207} (\bibinfo {year} {2003})}\BibitemShut
  {NoStop}%
\bibitem [{\citenamefont {Banerjee}\ and\ \citenamefont
  {Altman}(2016)}]{PhysRevLett.116.116601}%
  \BibitemOpen
  \bibfield  {author} {\bibinfo {author} {\bibfnamefont {S.}~\bibnamefont
  {Banerjee}}\ and\ \bibinfo {author} {\bibfnamefont {E.}~\bibnamefont
  {Altman}},\ }\href {\doibase 10.1103/PhysRevLett.116.116601} {\bibfield
  {journal} {\bibinfo  {journal} {Phys. Rev. Lett.}\ }\textbf {\bibinfo
  {volume} {116}},\ \bibinfo {pages} {116601} (\bibinfo {year}
  {2016})}\BibitemShut {NoStop}%
\bibitem [{\citenamefont {Iadecola}\ \emph {et~al.}(2015)\citenamefont
  {Iadecola}, \citenamefont {Santos},\ and\ \citenamefont
  {Chamon}}]{PhysRevB.92.125107}%
  \BibitemOpen
  \bibfield  {author} {\bibinfo {author} {\bibfnamefont {T.}~\bibnamefont
  {Iadecola}}, \bibinfo {author} {\bibfnamefont {L.~H.}\ \bibnamefont
  {Santos}}, \ and\ \bibinfo {author} {\bibfnamefont {C.}~\bibnamefont
  {Chamon}},\ }\href {\doibase 10.1103/PhysRevB.92.125107} {\bibfield
  {journal} {\bibinfo  {journal} {Phys. Rev. B}\ }\textbf {\bibinfo {volume}
  {92}},\ \bibinfo {pages} {125107} (\bibinfo {year} {2015})}\BibitemShut
  {NoStop}%
\bibitem [{\citenamefont {{von Keyserlingk}}\ and\ \citenamefont
  {{Sondhi}}(2016{\natexlab{a}})}]{2016arXiv160202157V}%
  \BibitemOpen
  \bibfield  {author} {\bibinfo {author} {\bibfnamefont {C.~W.}\ \bibnamefont
  {{von Keyserlingk}}}\ and\ \bibinfo {author} {\bibfnamefont {S.~L.}\
  \bibnamefont {{Sondhi}}},\ }\href@noop {} {\bibfield  {journal} {\bibinfo
  {journal} {ArXiv e-prints}\ } (\bibinfo {year} {2016}{\natexlab{a}})},\
  \Eprint {http://arxiv.org/abs/1602.02157} {arXiv:1602.02157
  [cond-mat.str-el]} \BibitemShut {NoStop}%
\bibitem [{\citenamefont {{von Keyserlingk}}\ and\ \citenamefont
  {{Sondhi}}(2016{\natexlab{b}})}]{2016arXiv160206949V}%
  \BibitemOpen
  \bibfield  {author} {\bibinfo {author} {\bibfnamefont {C.~W.}\ \bibnamefont
  {{von Keyserlingk}}}\ and\ \bibinfo {author} {\bibfnamefont {S.~L.}\
  \bibnamefont {{Sondhi}}},\ }\href@noop {} {\bibfield  {journal} {\bibinfo
  {journal} {ArXiv e-prints}\ } (\bibinfo {year} {2016}{\natexlab{b}})},\
  \Eprint {http://arxiv.org/abs/1602.06949} {arXiv:1602.06949
  [cond-mat.str-el]} \BibitemShut {NoStop}%
\bibitem [{\citenamefont {{Else}}\ and\ \citenamefont
  {{Nayak}}(2016)}]{2016arXiv160204804E}%
  \BibitemOpen
  \bibfield  {author} {\bibinfo {author} {\bibfnamefont {D.~V.}\ \bibnamefont
  {{Else}}}\ and\ \bibinfo {author} {\bibfnamefont {C.}~\bibnamefont
  {{Nayak}}},\ }\href@noop {} {\bibfield  {journal} {\bibinfo  {journal} {ArXiv
  e-prints}\ } (\bibinfo {year} {2016})},\ \Eprint
  {http://arxiv.org/abs/1602.04804} {arXiv:1602.04804 [cond-mat.str-el]}
  \BibitemShut {NoStop}%
\bibitem [{\citenamefont {Affleck}\ \emph {et~al.}(1987)\citenamefont
  {Affleck}, \citenamefont {Kennedy}, \citenamefont {Lieb},\ and\ \citenamefont
  {Tasaki}}]{PhysRevLett.59.799}%
  \BibitemOpen
  \bibfield  {author} {\bibinfo {author} {\bibfnamefont {I.}~\bibnamefont
  {Affleck}}, \bibinfo {author} {\bibfnamefont {T.}~\bibnamefont {Kennedy}},
  \bibinfo {author} {\bibfnamefont {E.~H.}\ \bibnamefont {Lieb}}, \ and\
  \bibinfo {author} {\bibfnamefont {H.}~\bibnamefont {Tasaki}},\ }\href
  {\doibase 10.1103/PhysRevLett.59.799} {\bibfield  {journal} {\bibinfo
  {journal} {Phys. Rev. Lett.}\ }\textbf {\bibinfo {volume} {59}},\ \bibinfo
  {pages} {799} (\bibinfo {year} {1987})}\BibitemShut {NoStop}%
\bibitem [{\citenamefont {Chen}\ and\ \citenamefont
  {Vishwanath}(2015)}]{PhysRevX.5.041034}%
  \BibitemOpen
  \bibfield  {author} {\bibinfo {author} {\bibfnamefont {X.}~\bibnamefont
  {Chen}}\ and\ \bibinfo {author} {\bibfnamefont {A.}~\bibnamefont
  {Vishwanath}},\ }\href {\doibase 10.1103/PhysRevX.5.041034} {\bibfield
  {journal} {\bibinfo  {journal} {Phys. Rev. X}\ }\textbf {\bibinfo {volume}
  {5}},\ \bibinfo {pages} {041034} (\bibinfo {year} {2015})}\BibitemShut
  {NoStop}%
\bibitem [{\citenamefont {Kitaev}(2009)}]{Kitaev10Fold}%
  \BibitemOpen
  \bibfield  {author} {\bibinfo {author} {\bibfnamefont {A.}~\bibnamefont
  {Kitaev}},\ }\href@noop {} {\bibfield  {journal} {\bibinfo  {journal} {AIP
  Conference Proceedings}\ }\textbf {\bibinfo {volume} {1134}},\ \bibinfo
  {pages} {22} (\bibinfo {year} {2009})}\BibitemShut {NoStop}%
\bibitem [{\citenamefont {Schnyder}\ \emph {et~al.}(2009)\citenamefont
  {Schnyder}, \citenamefont {Ryu}, \citenamefont {Furusaki},\ and\
  \citenamefont {Ludwig}}]{3149481}%
  \BibitemOpen
  \bibfield  {author} {\bibinfo {author} {\bibfnamefont {A.~P.}\ \bibnamefont
  {Schnyder}}, \bibinfo {author} {\bibfnamefont {S.}~\bibnamefont {Ryu}},
  \bibinfo {author} {\bibfnamefont {A.}~\bibnamefont {Furusaki}}, \ and\
  \bibinfo {author} {\bibfnamefont {A.~W.~W.}\ \bibnamefont {Ludwig}},\
  }\href@noop {} {\bibfield  {journal} {\bibinfo  {journal} {AIP Conference
  Proceedings}\ }\textbf {\bibinfo {volume} {1134}},\ \bibinfo {pages} {10}
  (\bibinfo {year} {2009})}\BibitemShut {NoStop}%
\bibitem [{\citenamefont {Schreiber}\ \emph {et~al.}(2015)\citenamefont
  {Schreiber}, \citenamefont {Hodgman}, \citenamefont {Bordia}, \citenamefont
  {L{\"u}schen}, \citenamefont {Fischer}, \citenamefont {Vosk}, \citenamefont
  {Altman}, \citenamefont {Schneider},\ and\ \citenamefont
  {Bloch}}]{Schreiber842}%
  \BibitemOpen
  \bibfield  {author} {\bibinfo {author} {\bibfnamefont {M.}~\bibnamefont
  {Schreiber}}, \bibinfo {author} {\bibfnamefont {S.~S.}\ \bibnamefont
  {Hodgman}}, \bibinfo {author} {\bibfnamefont {P.}~\bibnamefont {Bordia}},
  \bibinfo {author} {\bibfnamefont {H.~P.}\ \bibnamefont {L{\"u}schen}},
  \bibinfo {author} {\bibfnamefont {M.~H.}\ \bibnamefont {Fischer}}, \bibinfo
  {author} {\bibfnamefont {R.}~\bibnamefont {Vosk}}, \bibinfo {author}
  {\bibfnamefont {E.}~\bibnamefont {Altman}}, \bibinfo {author} {\bibfnamefont
  {U.}~\bibnamefont {Schneider}}, \ and\ \bibinfo {author} {\bibfnamefont
  {I.}~\bibnamefont {Bloch}},\ }\href {\doibase 10.1126/science.aaa7432}
  {\bibfield  {journal} {\bibinfo  {journal} {Science}\ }\textbf {\bibinfo
  {volume} {349}},\ \bibinfo {pages} {842} (\bibinfo {year}
  {2015})}\BibitemShut {NoStop}%
\bibitem [{\citenamefont {{Smith}}\ \emph {et~al.}(2015)\citenamefont
  {{Smith}}, \citenamefont {{Lee}}, \citenamefont {{Richerme}}, \citenamefont
  {{Neyenhuis}}, \citenamefont {{Hess}}, \citenamefont {{Hauke}}, \citenamefont
  {{Heyl}}, \citenamefont {{Huse}},\ and\ \citenamefont
  {{Monroe}}}]{2015arXiv150807026S}%
  \BibitemOpen
  \bibfield  {author} {\bibinfo {author} {\bibfnamefont {J.}~\bibnamefont
  {{Smith}}}, \bibinfo {author} {\bibfnamefont {A.}~\bibnamefont {{Lee}}},
  \bibinfo {author} {\bibfnamefont {P.}~\bibnamefont {{Richerme}}}, \bibinfo
  {author} {\bibfnamefont {B.}~\bibnamefont {{Neyenhuis}}}, \bibinfo {author}
  {\bibfnamefont {P.~W.}\ \bibnamefont {{Hess}}}, \bibinfo {author}
  {\bibfnamefont {P.}~\bibnamefont {{Hauke}}}, \bibinfo {author} {\bibfnamefont
  {M.}~\bibnamefont {{Heyl}}}, \bibinfo {author} {\bibfnamefont {D.~A.}\
  \bibnamefont {{Huse}}}, \ and\ \bibinfo {author} {\bibfnamefont
  {C.}~\bibnamefont {{Monroe}}},\ }\href@noop {} {\bibfield  {journal}
  {\bibinfo  {journal} {ArXiv e-prints}\ } (\bibinfo {year} {2015})},\ \Eprint
  {http://arxiv.org/abs/1508.07026} {arXiv:1508.07026 [quant-ph]} \BibitemShut
  {NoStop}%
\bibitem [{\citenamefont {Bordia}\ \emph {et~al.}(2016)\citenamefont {Bordia},
  \citenamefont {L\"uschen}, \citenamefont {Hodgman}, \citenamefont
  {Schreiber}, \citenamefont {Bloch},\ and\ \citenamefont
  {Schneider}}]{PhysRevLett.116.140401}%
  \BibitemOpen
  \bibfield  {author} {\bibinfo {author} {\bibfnamefont {P.}~\bibnamefont
  {Bordia}}, \bibinfo {author} {\bibfnamefont {H.~P.}\ \bibnamefont
  {L\"uschen}}, \bibinfo {author} {\bibfnamefont {S.~S.}\ \bibnamefont
  {Hodgman}}, \bibinfo {author} {\bibfnamefont {M.}~\bibnamefont {Schreiber}},
  \bibinfo {author} {\bibfnamefont {I.}~\bibnamefont {Bloch}}, \ and\ \bibinfo
  {author} {\bibfnamefont {U.}~\bibnamefont {Schneider}},\ }\href {\doibase
  10.1103/PhysRevLett.116.140401} {\bibfield  {journal} {\bibinfo  {journal}
  {Phys. Rev. Lett.}\ }\textbf {\bibinfo {volume} {116}},\ \bibinfo {pages}
  {140401} (\bibinfo {year} {2016})}\BibitemShut {NoStop}%
\bibitem [{\citenamefont {{Choi}}\ \emph {et~al.}(2016)\citenamefont {{Choi}},
  \citenamefont {{Hild}}, \citenamefont {{Zeiher}}, \citenamefont
  {{Schau{\ss}}}, \citenamefont {{Rubio-Abadal}}, \citenamefont {{Yefsah}},
  \citenamefont {{Khemani}}, \citenamefont {{Huse}}, \citenamefont {{Bloch}},\
  and\ \citenamefont {{Gross}}}]{2016arXiv160404178C}%
  \BibitemOpen
  \bibfield  {author} {\bibinfo {author} {\bibfnamefont {J.-y.}\ \bibnamefont
  {{Choi}}}, \bibinfo {author} {\bibfnamefont {S.}~\bibnamefont {{Hild}}},
  \bibinfo {author} {\bibfnamefont {J.}~\bibnamefont {{Zeiher}}}, \bibinfo
  {author} {\bibfnamefont {P.}~\bibnamefont {{Schau{\ss}}}}, \bibinfo {author}
  {\bibfnamefont {A.}~\bibnamefont {{Rubio-Abadal}}}, \bibinfo {author}
  {\bibfnamefont {T.}~\bibnamefont {{Yefsah}}}, \bibinfo {author}
  {\bibfnamefont {V.}~\bibnamefont {{Khemani}}}, \bibinfo {author}
  {\bibfnamefont {D.~A.}\ \bibnamefont {{Huse}}}, \bibinfo {author}
  {\bibfnamefont {I.}~\bibnamefont {{Bloch}}}, \ and\ \bibinfo {author}
  {\bibfnamefont {C.}~\bibnamefont {{Gross}}},\ }\href@noop {} {\bibfield
  {journal} {\bibinfo  {journal} {ArXiv e-prints}\ } (\bibinfo {year}
  {2016})},\ \Eprint {http://arxiv.org/abs/1604.04178} {arXiv:1604.04178
  [cond-mat.quant-gas]} \BibitemShut {NoStop}%
\bibitem [{\citenamefont {Lu}\ and\ \citenamefont
  {Vishwanath}(2016)}]{PhysRevB.93.155121}%
  \BibitemOpen
  \bibfield  {author} {\bibinfo {author} {\bibfnamefont {Y.-M.}\ \bibnamefont
  {Lu}}\ and\ \bibinfo {author} {\bibfnamefont {A.}~\bibnamefont
  {Vishwanath}},\ }\href {\doibase 10.1103/PhysRevB.93.155121} {\bibfield
  {journal} {\bibinfo  {journal} {Phys. Rev. B}\ }\textbf {\bibinfo {volume}
  {93}},\ \bibinfo {pages} {155121} (\bibinfo {year} {2016})}\BibitemShut
  {NoStop}%
\bibitem [{\citenamefont {Barkeshli}\ \emph {et~al.}(2014)\citenamefont
  {Barkeshli}, \citenamefont {Bonderson}, \citenamefont {Cheng},\ and\
  \citenamefont {Wang}}]{Barkeshli:2014cna}%
  \BibitemOpen
  \bibfield  {author} {\bibinfo {author} {\bibfnamefont {M.}~\bibnamefont
  {Barkeshli}}, \bibinfo {author} {\bibfnamefont {P.}~\bibnamefont
  {Bonderson}}, \bibinfo {author} {\bibfnamefont {M.}~\bibnamefont {Cheng}}, \
  and\ \bibinfo {author} {\bibfnamefont {Z.}~\bibnamefont {Wang}},\ }\href@noop
  {} {\  (\bibinfo {year} {2014})},\ \Eprint {http://arxiv.org/abs/1410.4540}
  {arXiv:1410.4540 [cond-mat.str-el]} \BibitemShut {NoStop}%
%%CITATION = ARXIV:1410.4540;%%
\bibitem [{\citenamefont {Fu}\ and\ \citenamefont
  {Kane}(2009)}]{PhysRevB.79.161408}%
  \BibitemOpen
  \bibfield  {author} {\bibinfo {author} {\bibfnamefont {L.}~\bibnamefont
  {Fu}}\ and\ \bibinfo {author} {\bibfnamefont {C.~L.}\ \bibnamefont {Kane}},\
  }\href {\doibase 10.1103/PhysRevB.79.161408} {\bibfield  {journal} {\bibinfo
  {journal} {Phys. Rev. B}\ }\textbf {\bibinfo {volume} {79}},\ \bibinfo
  {pages} {161408} (\bibinfo {year} {2009})}\BibitemShut {NoStop}%
\bibitem [{\citenamefont {Shivamoggi}\ \emph {et~al.}(2010)\citenamefont
  {Shivamoggi}, \citenamefont {Refael},\ and\ \citenamefont
  {Moore}}]{PhysRevB.82.041405}%
  \BibitemOpen
  \bibfield  {author} {\bibinfo {author} {\bibfnamefont {V.}~\bibnamefont
  {Shivamoggi}}, \bibinfo {author} {\bibfnamefont {G.}~\bibnamefont {Refael}},
  \ and\ \bibinfo {author} {\bibfnamefont {J.~E.}\ \bibnamefont {Moore}},\
  }\href {\doibase 10.1103/PhysRevB.82.041405} {\bibfield  {journal} {\bibinfo
  {journal} {Phys. Rev. B}\ }\textbf {\bibinfo {volume} {82}},\ \bibinfo
  {pages} {041405} (\bibinfo {year} {2010})}\BibitemShut {NoStop}%
\bibitem [{\citenamefont {Barkeshli}\ and\ \citenamefont
  {Qi}(2012)}]{PhysRevX.2.031013}%
  \BibitemOpen
  \bibfield  {author} {\bibinfo {author} {\bibfnamefont {M.}~\bibnamefont
  {Barkeshli}}\ and\ \bibinfo {author} {\bibfnamefont {X.-L.}\ \bibnamefont
  {Qi}},\ }\href {\doibase 10.1103/PhysRevX.2.031013} {\bibfield  {journal}
  {\bibinfo  {journal} {Phys. Rev. X}\ }\textbf {\bibinfo {volume} {2}},\
  \bibinfo {pages} {031013} (\bibinfo {year} {2012})}\BibitemShut {NoStop}%
\bibitem [{\citenamefont {Lindner}\ \emph {et~al.}(2012)\citenamefont
  {Lindner}, \citenamefont {Berg}, \citenamefont {Refael},\ and\ \citenamefont
  {Stern}}]{PhysRevX.2.041002}%
  \BibitemOpen
  \bibfield  {author} {\bibinfo {author} {\bibfnamefont {N.~H.}\ \bibnamefont
  {Lindner}}, \bibinfo {author} {\bibfnamefont {E.}~\bibnamefont {Berg}},
  \bibinfo {author} {\bibfnamefont {G.}~\bibnamefont {Refael}}, \ and\ \bibinfo
  {author} {\bibfnamefont {A.}~\bibnamefont {Stern}},\ }\href {\doibase
  10.1103/PhysRevX.2.041002} {\bibfield  {journal} {\bibinfo  {journal} {Phys.
  Rev. X}\ }\textbf {\bibinfo {volume} {2}},\ \bibinfo {pages} {041002}
  (\bibinfo {year} {2012})}\BibitemShut {NoStop}%
\bibitem [{\citenamefont {Fu}\ and\ \citenamefont
  {Kane}(2008)}]{PhysRevLett.100.096407}%
  \BibitemOpen
  \bibfield  {author} {\bibinfo {author} {\bibfnamefont {L.}~\bibnamefont
  {Fu}}\ and\ \bibinfo {author} {\bibfnamefont {C.~L.}\ \bibnamefont {Kane}},\
  }\href {\doibase 10.1103/PhysRevLett.100.096407} {\bibfield  {journal}
  {\bibinfo  {journal} {Phys. Rev. Lett.}\ }\textbf {\bibinfo {volume} {100}},\
  \bibinfo {pages} {096407} (\bibinfo {year} {2008})}\BibitemShut {NoStop}%
\bibitem [{\citenamefont {Cheng}(2012)}]{PhysRevB.86.195126}%
  \BibitemOpen
  \bibfield  {author} {\bibinfo {author} {\bibfnamefont {M.}~\bibnamefont
  {Cheng}},\ }\href {\doibase 10.1103/PhysRevB.86.195126} {\bibfield  {journal}
  {\bibinfo  {journal} {Phys. Rev. B}\ }\textbf {\bibinfo {volume} {86}},\
  \bibinfo {pages} {195126} (\bibinfo {year} {2012})}\BibitemShut {NoStop}%
\bibitem [{\citenamefont {Clarke}\ \emph {et~al.}(2013)\citenamefont {Clarke},
  \citenamefont {Alicea},\ and\ \citenamefont {Shtengel}}]{CAKpara}%
  \BibitemOpen
  \bibfield  {author} {\bibinfo {author} {\bibfnamefont {D.~J.}\ \bibnamefont
  {Clarke}}, \bibinfo {author} {\bibfnamefont {J.}~\bibnamefont {Alicea}}, \
  and\ \bibinfo {author} {\bibfnamefont {K.}~\bibnamefont {Shtengel}},\ }\href
  {http://dx.doi.org/10.1038/ncomms2340} {\bibfield  {journal} {\bibinfo
  {journal} {Nat Commun}\ }\textbf {\bibinfo {volume} {4}},\ \bibinfo {pages}
  {1348} (\bibinfo {year} {2013})}\BibitemShut {NoStop}%
\bibitem [{\citenamefont {Mong}\ \emph {et~al.}(2014)\citenamefont {Mong},
  \citenamefont {Clarke}, \citenamefont {Alicea}, \citenamefont {Lindner},
  \citenamefont {Fendley}, \citenamefont {Nayak}, \citenamefont {Oreg},
  \citenamefont {Stern}, \citenamefont {Berg}, \citenamefont {Shtengel},\ and\
  \citenamefont {Fisher}}]{PhysRevX.4.011036}%
  \BibitemOpen
  \bibfield  {author} {\bibinfo {author} {\bibfnamefont {R.~S.~K.}\
  \bibnamefont {Mong}}, \bibinfo {author} {\bibfnamefont {D.~J.}\ \bibnamefont
  {Clarke}}, \bibinfo {author} {\bibfnamefont {J.}~\bibnamefont {Alicea}},
  \bibinfo {author} {\bibfnamefont {N.~H.}\ \bibnamefont {Lindner}}, \bibinfo
  {author} {\bibfnamefont {P.}~\bibnamefont {Fendley}}, \bibinfo {author}
  {\bibfnamefont {C.}~\bibnamefont {Nayak}}, \bibinfo {author} {\bibfnamefont
  {Y.}~\bibnamefont {Oreg}}, \bibinfo {author} {\bibfnamefont {A.}~\bibnamefont
  {Stern}}, \bibinfo {author} {\bibfnamefont {E.}~\bibnamefont {Berg}},
  \bibinfo {author} {\bibfnamefont {K.}~\bibnamefont {Shtengel}}, \ and\
  \bibinfo {author} {\bibfnamefont {M.~P.~A.}\ \bibnamefont {Fisher}},\ }\href
  {\doibase 10.1103/PhysRevX.4.011036} {\bibfield  {journal} {\bibinfo
  {journal} {Phys. Rev. X}\ }\textbf {\bibinfo {volume} {4}},\ \bibinfo {pages}
  {011036} (\bibinfo {year} {2014})}\BibitemShut {NoStop}%
\bibitem [{\citenamefont {Fendley}(2012)}]{1742-5468-2012-11-P11020}%
  \BibitemOpen
  \bibfield  {author} {\bibinfo {author} {\bibfnamefont {P.}~\bibnamefont
  {Fendley}},\ }\href@noop {} {\bibfield  {journal} {\bibinfo  {journal}
  {Journal of Statistical Mechanics: Theory and Experiment}\ }\textbf {\bibinfo
  {volume} {2012}},\ \bibinfo {pages} {P11020} (\bibinfo {year}
  {2012})}\BibitemShut {NoStop}%
\bibitem [{\citenamefont {Alicea}\ and\ \citenamefont
  {Fendley}(2016)}]{AFpara}%
  \BibitemOpen
  \bibfield  {author} {\bibinfo {author} {\bibfnamefont {J.}~\bibnamefont
  {Alicea}}\ and\ \bibinfo {author} {\bibfnamefont {P.}~\bibnamefont
  {Fendley}},\ }\href@noop {} {\bibfield  {journal} {\bibinfo  {journal}
  {Annual Review of Condensed Matter Physics}\ }\textbf {\bibinfo {volume}
  {7}},\ \bibinfo {pages} {119} (\bibinfo {year} {2016})}\BibitemShut {NoStop}%
\bibitem [{\citenamefont {Nayak}\ \emph {et~al.}(2008)\citenamefont {Nayak},
  \citenamefont {Simon}, \citenamefont {Stern}, \citenamefont {Freedman},\ and\
  \citenamefont {Das~Sarma}}]{RevModPhys.80.1083}%
  \BibitemOpen
  \bibfield  {author} {\bibinfo {author} {\bibfnamefont {C.}~\bibnamefont
  {Nayak}}, \bibinfo {author} {\bibfnamefont {S.~H.}\ \bibnamefont {Simon}},
  \bibinfo {author} {\bibfnamefont {A.}~\bibnamefont {Stern}}, \bibinfo
  {author} {\bibfnamefont {M.}~\bibnamefont {Freedman}}, \ and\ \bibinfo
  {author} {\bibfnamefont {S.}~\bibnamefont {Das~Sarma}},\ }\href {\doibase
  10.1103/RevModPhys.80.1083} {\bibfield  {journal} {\bibinfo  {journal} {Rev.
  Mod. Phys.}\ }\textbf {\bibinfo {volume} {80}},\ \bibinfo {pages} {1083}
  (\bibinfo {year} {2008})}\BibitemShut {NoStop}%
\bibitem [{\citenamefont {Feiguin}\ \emph {et~al.}(2007)\citenamefont
  {Feiguin}, \citenamefont {Trebst}, \citenamefont {Ludwig}, \citenamefont
  {Troyer}, \citenamefont {Kitaev}, \citenamefont {Wang},\ and\ \citenamefont
  {Freedman}}]{PhysRevLett.98.160409}%
  \BibitemOpen
  \bibfield  {author} {\bibinfo {author} {\bibfnamefont {A.}~\bibnamefont
  {Feiguin}}, \bibinfo {author} {\bibfnamefont {S.}~\bibnamefont {Trebst}},
  \bibinfo {author} {\bibfnamefont {A.~W.~W.}\ \bibnamefont {Ludwig}}, \bibinfo
  {author} {\bibfnamefont {M.}~\bibnamefont {Troyer}}, \bibinfo {author}
  {\bibfnamefont {A.}~\bibnamefont {Kitaev}}, \bibinfo {author} {\bibfnamefont
  {Z.}~\bibnamefont {Wang}}, \ and\ \bibinfo {author} {\bibfnamefont {M.~H.}\
  \bibnamefont {Freedman}},\ }\href {\doibase 10.1103/PhysRevLett.98.160409}
  {\bibfield  {journal} {\bibinfo  {journal} {Phys. Rev. Lett.}\ }\textbf
  {\bibinfo {volume} {98}},\ \bibinfo {pages} {160409} (\bibinfo {year}
  {2007})}\BibitemShut {NoStop}%
\bibitem [{\citenamefont {Trebst}\ \emph {et~al.}(2008)\citenamefont {Trebst},
  \citenamefont {Ardonne}, \citenamefont {Feiguin}, \citenamefont {Huse},
  \citenamefont {Ludwig},\ and\ \citenamefont
  {Troyer}}]{PhysRevLett.101.050401}%
  \BibitemOpen
  \bibfield  {author} {\bibinfo {author} {\bibfnamefont {S.}~\bibnamefont
  {Trebst}}, \bibinfo {author} {\bibfnamefont {E.}~\bibnamefont {Ardonne}},
  \bibinfo {author} {\bibfnamefont {A.}~\bibnamefont {Feiguin}}, \bibinfo
  {author} {\bibfnamefont {D.~A.}\ \bibnamefont {Huse}}, \bibinfo {author}
  {\bibfnamefont {A.~W.~W.}\ \bibnamefont {Ludwig}}, \ and\ \bibinfo {author}
  {\bibfnamefont {M.}~\bibnamefont {Troyer}},\ }\href {\doibase
  10.1103/PhysRevLett.101.050401} {\bibfield  {journal} {\bibinfo  {journal}
  {Phys. Rev. Lett.}\ }\textbf {\bibinfo {volume} {101}},\ \bibinfo {pages}
  {050401} (\bibinfo {year} {2008})}\BibitemShut {NoStop}%
\bibitem [{\citenamefont {Santos}(2015)}]{PhysRevB.91.155150}%
  \BibitemOpen
  \bibfield  {author} {\bibinfo {author} {\bibfnamefont {L.~H.}\ \bibnamefont
  {Santos}},\ }\href {\doibase 10.1103/PhysRevB.91.155150} {\bibfield
  {journal} {\bibinfo  {journal} {Phys. Rev. B}\ }\textbf {\bibinfo {volume}
  {91}},\ \bibinfo {pages} {155150} (\bibinfo {year} {2015})}\BibitemShut
  {NoStop}%
\end{thebibliography}%

\appendix

\section{Finite-depth unitary mapping from SPT to paramagnet -- an example \label{app:SPT}}
\label{AppA}

In this section, we explicitly construct an example of a finite depth unitary transformation that converts the SPT to a trivial paramagnet. We consider a discrete version of the Haldane spin chain with symmetry group $\Z_2\times\Z_2$. The chain contains spin-1/2 degrees of freedom, and has two-sublattices (even and odd sites). The zero-correlation length paramagnet reads:
\begin{align}
H_\text{PM} = -\sum_{i=1}^{2L} h_i\sigma^x_i.
\end{align}
The symmetry generators are $g_e = \prod_{i}\sigma^x_{2i}$, $g_o = \prod_i\sigma^x_{2i+1}$, which flip the spins about the z-axis on the even and odd sublattice respectively.

The SPT phase is described by zero-correlation length Hamiltonian:
\begin{align}
H_\text{SPT} = -\sum_{i=2}^{2L-1} J_i\sigma^z_{i-1}\sigma^x_i\sigma^z_{i+1}.
\end{align}
Since $H_\text{SPT}$ is a non-trivial SPT phase, then by definition, we cannot continuously deform the eigenstates of $H_\text{SPT}$ to those of the trivial paramagnet $H_\text{PM}$. Namely, there is no continuous family of symmetry-preserving unitary operators $U(\lambda)$ for $\lambda\in[0,1]$, where $U(0)=1$ and $U(\lambda=1) = U_\text{SPT}$, where $U_\text{SPT}$ maps the SPT states to trivial ones. 

However, if we sacrifice the continuity, we can write down the end result, $U(\lambda=1)$, as a finite-depth unitary circuit, which is all that is required to establish the local representation of symmetry. An explicit construction that does the job is
\begin{align}
U_\text{SPT} = e^{-i\pi/4\sum_i \sigma^z_i\sigma^z_{i+1}},
\end{align}
which takes $\sigma^x_i \mapsto \sigma^z_{i-1}\sigma^x_{i}\sigma^z_{i+1}$ everywhere in the bulk of the chain (see {\it eg.}~\onlinecite{PhysRevB.91.155150}). The unitary operator $U_\text{SPT}$ is finite depth, and preserves the form of the symmetry generators $g_e$ and $g_o$, except at the boundaries of the system. 

%This construction does not invalidate the claimed SPT nature of the eigenstates of $H_\text{SPT}$, since, if we attempt to define $U_\text{SPT}$ as the end result of a continuous deformation generated by the family of unitaries $U_\text{SPT}(\lambda)\equiv e^{-i\lambda A}$, with $\lambda\in [0,1]$, which continuously maps the SPT phase ($\lambda=0$) to the trivial one ($\lambda=1$), then we see that this continuous deformation cannot be done in a symmetry preserving manner (i.e. $A=\sum_i\sigma^z_i\sigma^z_{i+1}$ does not commute with the symmetry generators $g_e$ or $g_o$). 

While we have constructed an explicit example of the desired finite-depth unitary $U_\text{SPT}$ for this particular symmetry class, similar constructions can be made for any general SPT class~\cite{PhysRevB.87.155114}.

\end{document}